\documentclass[%
 aip,
 amsmath,amssymb,
preprint,%
]{revtex4-1}

\usepackage{graphicx}
\usepackage{natbib}
\usepackage{color}
\usepackage{todonotes}
\usepackage{latexsym}
\usepackage{natbib}
\usepackage{upgreek}
\usepackage{mathrsfs}
\usepackage{float}
\usepackage{caption}
\usepackage{soul}
\usepackage{textcomp}
\usepackage{stmaryrd}
\usepackage{amsmath}
\usepackage{mathtools}

\usepackage{subcaption}

\def\ii{{\rm i}}
\DeclareMathOperator{\sech}{sech}

\begin{document}

\preprint{AIP/123-QED}

\title{On the inertial effects of density variation in stratified shear flows}
\author{Anirban Guha}
\email{anirbanguha.ubc@gmail.com}

\author{Raunak Raj}
\affiliation{
Environmental and Geophysical Fluids Group, Department of Mechanical Engineering, Indian Institute of Technology Kanpur, U.P. 208016, India.\\}%


\date{\today}

\begin{abstract}
In this paper, we first revisit the celebrated Boussinesq approximation in stratified  flows. Using scaling arguments we show that when the background shear is weak, {the} Boussinesq approximation yields {either (i)} $A_t\ll \mathcal{O}(1)$ or (ii) $Fr_c^2 \ll \mathcal{O}(1)$, where $A_t$ is the ratio of density variation to the mean density and  $Fr_c$ is the ratio of the phase speed to the long wave speed. The second clause implies, contrary to the commonly accepted notion, that a flow with large density variations can also be Boussinesq. Indeed, we show that deep water surface gravity waves are Boussinesq while shallow water surface gravity  waves aren’t. 
However, in the presence of moderate/strong shear,  Boussinesq approximation implies the conventionally accepted   $A_t\ll \mathcal{O}(1)$.
To understand the inertial effects of density variation, {our second objective is to} explore various non-Boussinesq shear flows and study  different kinds of {stably propagating waves} that can be present at an interface between two fluids of different background densities and vorticities. Furthermore, three kinds of density interfaces -- neutral, stable and unstable -- embedded in a background shear layer,  are investigated. Instabilities ensuing from these configurations, which includes Kelvin-Helmholtz, Holmboe, Rayleigh-Taylor and triangular-jet, are studied in terms of resonant wave interactions. {The effects of density stratification and the shear on the stability of each of these flow configurations are explored.} Some of the results, e.g.\ the destabilizing role of density stratification, stabilizing role of shear, etc.\ are apparently counter-intuitive, but physical explanations are possible if the instabilities are interpreted from wave interactions perspective.  
\end{abstract}


\maketitle

\section{Introduction}
Hydrodynamic instabilities occurring in parallel flows have numerous industrial and environmental applications, and have therefore  been comprehensively investigated in the past \citep{drazin2004hydrodynamic,charru2011hydrodynamic}. Parallel flows are often produced when two streams of different velocities and densities flow side by side.  For example in  fuel injection systems, a fast stream of gas flows over a slow stream of liquid (fuel), making the liquid--gas interface prone to shear instabilities \citep{marmottant_villermaux_2004,behzad2014role}. Here density stratification plays a crucial role in the instability dynamics, however \emph{gravity does not}. {Such parallel flow configurations} and ensuing instabilities are not restricted to industrial two-phase flows, {and} are {also} observed in systems that are fundamentally different, {such as}\  coronal mass ejections \citep{foullon2011magnetic,ofman2011sdo}, magnetosphere--solar-wind interactions \citep{nykyri2017influence}, {among several other applications.}
In a variety of flows occurring in the geophysical context (especially in oceans, lakes, estuaries), density {variation} is often found to be small. This small density  {variation} has an insignificant effect on the inertial acceleration of the fluid parcels, but has a non-trivial effect on the buoyancy force (meaning, \emph{gravity does} play a major role here). 
This led to the famous Boussinesq approximation \citep{turner1979buoyancy},
which allows, for example, to successfully capture  internal gravity waves without dealing with other complexities associated with  density variations in the governing Navier--Stokes equations. Since {the} Boussinesq approximation is widely used in density stratified flows, {by density variation we often} inadvertently presume buoyancy variation.
Boussinesq approximation is often interpreted as a technique that filters out the fast (sound) waves arising from compressibility effects (large density variations because of pressure), but keeps slower waves that exist in an incompressible flow due to small density variations  (e.g., density variations in water because of salinity or temperature) \citep{suth2010}. 
As mentioned previously, many  multi-phase systems {are incompressible flows having} fluids of different densities (not necessarily small) flowing side by side, e.g.\ air-sea interface \citep{shete2017effect}, fuel injectors \citep{marmottant_villermaux_2004,behzad2014role},  displacement flows \citep{ali2018displacement,hasnain2017buoyant}. Finite density variations in the same fluid due to a stratifying agent (like temperature) is also possible, e.g.\ atmospheric flows.
In this paper, any incompressible, density stratified {flow} in which density stratification has a non-trivial effect on the inertial acceleration will be termed  \emph{non-Boussinesq}. In such flows, Boussinesq approximation can be an over-simplification since it not only filters out the sound waves, it  also eliminates all kinds of non-Boussinesq waves from the system.

The primary objective of this paper is to fundamentally understand the importance of density variation in the inertial terms, especially in the presence of background shear. 
Since {the} vorticity evolution equation fundamentally governs {the} density (and {the} buoyancy) stratified shear flows, we revisit this equation to identify the role of each vorticity generating term, especially when {the} Boussinesq approximation is not imposed \emph{a-priori}. This, we believe, can underpin the actual meaning of {the} Boussinesq approximation, and its true implications. 

The interplay/competition between shear and buoyancy drives the ensuing instabilities  in Boussinesq shear flows, and a physical understanding of this can be obtained by studying the interaction between the waves present at the vorticity and buoyancy interfaces \cite[]{holmboe1962behavior,sakai1989rossby,baines1994mechanism,caulfield1994multiple,heifetz1999counter,heif2005,carp2012,guha2014wave}. The  wave interaction perspective not only provides a mechanistic understanding of the existing shear instabilities,  often it  allows one to make  \emph{a-priori} judgments regarding the likelihood of a given stratified shear flow set-up to {become} unstable, and if so, which waves (effects) would play dominant roles; see for example \citet{shete2017effect}.  
A wave present at a vorticity interface is called a ``vorticity wave'', while those present at a buoyancy interface are called ``interfacial gravity waves''. While the interfacial waves present in Boussinesq shear flows are well studied, those in non-Boussinesq shear flows, {are a recent development}. 
There is a good possibility for these waves to interact between themselves, or with other wave(s) existing in the system, to yield different kinds of non-Boussinesq shear instabilities. {In this paper, we study} the non-Boussinesq versions of Kelvin-Helmholtz instability, Holmboe instability, Rayleigh-Taylor instability as well as the instability of a triangular jet \citep{drazin2002introduction} and interpreted each of them in terms of wave interactions. 

The paper is organized as follows. In Sec. \ref{sec:gov_eq} we scrutinize the vorticity evolution equation in order to understand the meaning as well as the validity of the celebrated Boussinesq approximation. Contrary to the common notion, we show that {the} Boussinesq approximation can be valid even when the density stratification is large. Next we focus on a single interface between two fluids\cite{drazin2004hydrodynamic,suth2010} of different background densities and vorticities to obtain the most general dispersion relation of linear interfacial waves {in order to obtain the growth rate of instabilities and phase speed of stable waves}. Possibility of the existence of non-Boussinesq waves and their generation mechanisms are discussed.  In Sec. \ref{sec:tr_jet}, we consider a neutrally stratified flow {(no gravity)} configuration - a triangular jet of constant density flowing through a quiescent medium of a different (constant) density. The instability mechanism is understood in terms of lesser known non-Boussinesq interfacial waves.  In Sec. \ref{sec:KH}, we study non-Boussinesq Kelvin-Helmholtz instability for neutral stratification and non-Boussinesq Holmboe instability for stable stratification. In Sec. \ref{sec:RT_111}, the effect of shear on unstably stratified density interface is studied (which is basically Rayleigh-Taylor instability in a shear layer). The paper is summarized and concluded in Sec. \ref{sec:Sum_con}.

\section{Governing equations}
\label{sec:gov_eq}
The governing incompressible continuity and (inviscid) Navier-Stokes equations of a density stratified  fluid are given below:
\renewcommand{\theequation}{\arabic{section}.\arabic{equation}a,b}
\begin{equation}
\nabla \cdot \mathbf{u}=0,\,\,\, \frac{D\rho}{Dt}=0,
\label{eq:gov_NSC_1}
\end{equation} 
\addtocounter{equation}{-1}
\renewcommand{\theequation}{\arabic{section}.\arabic{equation}c}
\begin{equation} 
\frac{D\mathbf{u}}{Dt}=-g\hat{\mathbf{k}}-\frac{1}{\rho}\nabla p.
\label{eq:gov_NSC_2}
\end{equation} 
Here $D/Dt \equiv \partial/\partial t + \mathbf{u}\cdot \nabla$ is the material derivative, and $\mathbf{u}$, $\rho$, $p$ and $g$ respectively denote velocity, density, pressure and acceleration due to gravity. {$\mathbf{\hat{k}}$ is a unit vector pointing in a direction opposite to gravity }. The vorticity equation is obtained by taking curl of Eq.\ (\ref{eq:gov_NSC_2}) and using Eq.\ (\ref{eq:gov_NSC_1}):
\renewcommand{\theequation}{\arabic{section}.\arabic{equation}}
\begin{equation}
\frac{D\boldsymbol{q}}{Dt}=\underbrace{ (\boldsymbol{q} \cdot \nabla)\mathbf{u}}_\text{\clap{Stretching}}  +\underbrace{\frac{1}{\rho^2}\nabla\rho\times\nabla p}_\text{\clap{Baroclinic}},
\label{eq:full-vort}
 \end{equation}
 where $\boldsymbol{q}\equiv \nabla \times \mathbf{u}$ is the vorticity. The first term on the RHS is the vortex stretching term that would be absent in 2D flows. The second term denotes the baroclinic generation of vorticity i.e.  vorticity generated when the isopycnals and isobars are not parallel. In general, this term can be divided into two parts -  ``gravitational'' and  ``non-Boussinesq''.

We consider a 2D flow $\mathbf{u}=(u,\,w)$ in the $x-z$ plane, and assume a base state that varies only along $z$: $\bar{u}=\bar{u}(z)$, $\bar{w}=0$, $\bar{q}(z)=d\bar{u}/dz$, $\bar{p}=\bar{p}(z)$ and $\bar{\rho}=\bar{\rho}(z)$. The base state follows hydrostatic pressure balance $d\bar{p}/dz=-\bar{\rho}g$. We add perturbations to the base flow and the governing equations are linearized. This produces the perturbation vorticity evolution equation
\begin{equation}\label{eq:2.3}
\frac{\bar{D}q}{\bar{D}t}=\,\,\,\, \underbrace{- {w}\frac{d\bar{q}}{dz}}_\text{\clap{$T_1\equiv$Barotropic}}\,\,\,\,\,\,\,\,\,\,\,\,\,\,\,\,\,+\,\,\,\,\,\,\,\,\,\,\,\,  \underbrace{\frac{g}{\bar{\rho}}\frac{\partial  {\rho}}{\partial x}}_\text{\clap{$T_2\equiv$Gravitational baroclinic}}\,\,\,\,\,\,\,\,\,\,\,\,\,\,\,\,\,\,\,\,\,\,\,\,+\,\,\,\,\,\,\,\,\,\,\,\,\,\,\,\,\,\, \underbrace{\frac{1}{\bar{\rho}^{2}}\Big(\frac{d \bar{\rho}}{d z}\frac{\partial  {p}}{\partial x} \Big)}_\text{\clap{$T_3\equiv$Non-Boussinesq baroclinic}},
\end{equation}
where  $\bar{D}/\bar{D}t\equiv \partial/\partial t + \bar{u}\, \partial/\partial x$ denotes the \emph{linearized} material derivative, and the quantities with no overbars are the perturbation quantities (we note here that `total variables' in Eq.\ (\ref{eq:gov_NSC_1})--(\ref{eq:full-vort}) have also been referred to as quantities with no overbars).  Eq.\ (\ref{eq:2.3}) shows that vorticity can be generated from three sources:\\
(i) $T_1\equiv-w {d\bar{q}}/{dz}$: This term  denotes the barotropic generation of vorticity, i.e.\ vorticity generated due to gradients in background vorticity. For example, if we are considering rotating flows in the $\beta$-plane, then  $d\bar{q}/dz$  basically becomes the planetary vorticity gradient $\beta$. 
\\(ii) $T_2 \equiv (g/\bar{\rho}){\partial \rho}/{\partial x}$: This term denotes the \textcolor{black}{``gravitational baroclinic torque''}, and is responsible for the propagation of interfacial gravity waves at the pycnocline. In fact, it is the only baroclinic generation term present when the Boussinesq approximation is invoked. We also note here that because of  this term, density variation often becomes synonymous with buoyancy variation. In absence of gravity, this term would vanish identically.
\\(iii) $T_3\equiv (1/\bar{\rho}^{2})(d \bar{\rho}/{d z})({\partial p}/{\partial x})$: This term, which denotes the ``non-Boussinesq baroclinic generation of vorticity'', arises out of the density variations in the inertial terms, and is completely independent of the gravitational effects. This term vanishes if Boussinesq approximation (which, in a simplified sense, implies $\bar{\rho} \approx \mathrm{const.}$ in Eq.\ (\ref{eq:2.3})) is invoked.

\subsection{Boussinesq approximation in shear flows and its validity}
\label{sec:Bouss_val}
{The validity of the  Boussinesq approximation in shear flows} stems from Eq.\ (\ref{eq:2.3}) -- it  implies that the non-Boussinesq baroclinic term $T_3$ is negligible in comparison to either  the barotropic term $T_1$ or the gravitational baroclinic term $T_2$,  i.e., 
\begin{equation}
\mathcal{O}\underbrace{\left(-w\frac{d\bar{q}}{dz}\right)}_\text{\clap{$T_1$}}\gg \mathcal{O}\underbrace{\left(\frac{1}{\bar{\rho}^2}\frac{d \bar{\rho}}{d z}\frac{\partial  {p}}{\partial x} \right)}_\text{\clap{$T_3$}}\,\,\,\,\,\mathrm{or}\,\,\,\,\mathcal{O}\underbrace{\left(\frac{g}{\bar{\rho}}\frac{\partial  {\rho}}{\partial x}\right)}_\text{\clap{$T_2$}}\gg \mathcal{O}\underbrace{\left(\frac{1}{\bar{\rho}^2}\frac{d \bar{\rho}}{d z}\frac{\partial  {p}}{\partial x} \right)}_\text{\clap{$T_3$}}.
\end{equation}
To make sense of the above conditions, especially to cast them in terms of the non-dimensional numbers,  we use the linearized versions of Eq.\ (\ref{eq:gov_NSC_1}) and the $x$-component of Eq.\ (\ref{eq:gov_NSC_2}), keeping in mind that quantities without overbars are perturbation quantities:
\begin{equation}
\renewcommand{\theequation}{\arabic{section}.\arabic{equation}a,b}
 \frac{\partial u}{\partial x}+\frac{\partial w}{\partial z}=0,\,\,\,\frac{\bar{D}\rho}{\bar{D}t}\equiv\frac{\partial \rho}{\partial t}+ \bar{u}\frac{\partial \rho}{\partial x}=-w\frac{d\bar{\rho}}{dz},
\label{eq:gov_NSC_11}
\end{equation} 
\addtocounter{equation}{-1}
\renewcommand{\theequation}{\arabic{section}.\arabic{equation}c}
\begin{equation} 
\frac{\bar{D}u}{\bar{D} t}+w\bar{q}\equiv \frac{\partial u}{\partial t}+ \bar{u}\frac{\partial u}{\partial x}+ w\bar{q}= -\frac{1}{\bar{\rho}}\frac{\partial p}{\partial x}.
\label{eq:gov_NSC_12}
\end{equation} 
Let {$L\equiv$ horizontal length scale}, $\Delta U\equiv$ shear velocity scale, { $W \equiv$ vertical velocity scale}, $H \equiv$ shear length scale, $\chi \equiv$ length scale for the variation of $\bar{\rho}$, $\Delta \rho \equiv$ scale denoting  variation of $\bar{\rho}$, { $ \rho^{*}\equiv$ scale denoting perturbation density, $\Delta P \equiv$  pressure scale}, and $\rho_0 \equiv$ reference density. We take $1/\omega$ as the timescale, where $\omega$ is a characteristic gravity wave frequency. 
Scaling analysis of Eq.\ (\ref{eq:gov_NSC_11}a) yields 
$$ W \sim \frac{\Delta U H}{L},$$
while that of Eq.\ (\ref{eq:gov_NSC_11}b) yields 
$$\max\left[\mathcal{O}\left(\rho^{*}\omega\right),\mathcal{O}\left(\frac{\rho^{*}\Delta U}{L}\right)\right]=\mathcal{O}\left(\frac{\Delta U\Delta \rho H}{\chi L}\right).$$
Similarly, scaling analysis of  Eq.\ (\ref{eq:gov_NSC_12}) shows
$$\max\left[\mathcal{O}\left(\Delta U\omega\right),\mathcal{O}\left(\frac{\Delta U^2}{L}\right)\right]=\mathcal{O}\left(\frac{\Delta P}{\rho_0 L}\right).$$
We define a few non-dimensional variables: (i) Atwood number, $A_t \equiv \Delta\rho/ (2\rho_0)$, (ii) \emph{shear} Froude number, $Fr \equiv \Delta U/\sqrt{gH}$, and (iii)  a non-conventional Froude number based on the ratio of the phase speed $c$ to the long wave speed,  $Fr_c\equiv c/\sqrt{gH}$, where $c=L\omega$. 

\subsubsection{Boussinesq approximation in moderate/strongly sheared flows:  $Fr \gtrsim \mathcal{O}(1)$}
\label{sec:BAV}
When shear is moderate or large, we have  $$\frac{\Delta U}{L} \gtrsim
 \omega\,\,\,\, \implies \,\,\,\, Fr \gtrsim
 Fr_c.$$ 
In this situation, the convective term of the material derivatives in Eq.\ (\ref{eq:gov_NSC_11}b)-Eq.\ (\ref{eq:gov_NSC_12}) provide the dominant balance. Hence
$$\rho^{*}\sim\frac{\Delta \rho H}{\chi}\,\,\,\, \mathrm{and} \,\,\,\, \Delta P \sim \rho_0 \Delta U^2.$$
Thus we obtain

$$T_1=\mathcal{O}\left(\frac{\Delta U^2}{H L}\right),\,\, 
T_2= \mathcal{O}\left(\frac{g \Delta \rho H}{\rho_0 \chi L}\right),\,\, \mathrm{and}\,\, T_3= \mathcal{O}\left( \frac{\Delta \rho \Delta U^2}{\rho_0 \chi L}\right).$$ 
Therefore, Boussinesq approximation in moderate to strongly sheared flows imply
\renewcommand{\theequation}{\arabic{section}.\arabic{equation}}
\begin{align}
\mathrm{either}\,\,\,\,\,\,\frac{T_3}{T_1}= \mathcal{O}\left(A_t \frac{H}{\chi}\right) \ll  \mathcal{O}(1) \implies A_t \ll \mathcal{O}\left( \frac{\chi}{H}\right),
\label{eq:Bouss_app1}
\end{align}
\begin{align}
\mathrm{or}\,\,\,\,\,\,\frac{T_3}{T_2}\sim Fr^{2} \ll  \mathcal{O}(1),
\label{eq:Bouss_app2}
\end{align}
 The ratio between base density thickness  and base shear thickness, $\chi/H$, will depend on the stratified shear flow situation one is interested in. These thicknesses can be determined by the diffusivities of the corresponding agents:  $H$ is determined by the {velocity scale and the} diffusion of momentum (coefficient of viscosity), while $\chi$ in oceans is determined by the {velocity scale and the} diffusion of heat or salt. {It is imperative to understand here that the base flow can be viscous and diffusive, however the instability is inviscid and non-diffusive}. \citet{smyth1988finite} have shown that
\begin{equation*}
\frac{\chi}{H} \approx Pr^{-1/2},
\end{equation*}
where $Pr$ is the Prandtl number. Typical values of $Pr$  in geophysical flows are\cite{rahmani2011kelvin}  -- (a) $Pr\sim 1$ for atmosphere (thermal stratification) (b)  $Pr \sim 9$ for lakes (thermal stratification), (c)  $Pr \sim 700$ for oceans (salt stratification). Hence, at least in geophysical context, this implies
\begin{equation*}
\frac{\chi}{H} \leq 1,
\end{equation*}
and hence, the Boussinesq approximation here  can be expressed as follows:
$$ \mathrm{either}\,\,\, A_t \ll \mathcal{O}\left(1\right)\,\, \mathrm{or}\,\, Fr^{2}   \ll  \mathcal{O}(1).$$
 Since we have \emph{a-priori} assumed $Fr\gtrsim O(1)$ for moderate/large stratification, that is, the shear is not small, the condition  $Fr^2 \ll \mathcal{O}(1)$ is inconsistent with our assumption. Therefore, for moderate to  strongly sheared flows, we have 
 \begin{equation}
A_t \ll \mathcal{O}(1).
\label{eq:Bouss_app}
\end{equation}
Thus ``small density variations'' is the \emph{only} condition for the validity of the Boussinesq approximation in density stratified flows where shear is significant. This finding is along the lines of the existing understanding of ``Boussinesqness'', see \citet{suth2010,spiegel1960boussinesq,turner1979buoyancy}.

{
\subsubsection{Boussinesq approximation when the shear is weak:  $Fr \ll \mathcal{O}(1)$}
\label{sec:SGW_bnb}
When shear is  weak, the timescale is determined by the  gravity wave's timeperiod:  $$\frac{\Delta U}{L} \ll \omega\,\,\,\, \implies \,\,\,\, Fr\ll Fr_c.$$
In this situation, the temporal term of the material derivatives in Eq.\ (\ref{eq:gov_NSC_11}b)-Eq.\ (\ref{eq:gov_NSC_12}) provide the dominant balance. Scaling analysis yields
$$T_1= \mathcal{O}\left(\frac{\Delta U^2}{H L}\right),\,\, 
T_2= \mathcal{O}\left(\frac{g \Delta \rho \Delta U H}{\rho_0 \omega \chi L^2}\right),\,\, \mathrm{and}\,\, T_3= \mathcal{O}\left( \frac{\Delta \rho \Delta U \omega}{\rho_0 \chi}\right).$$ 
As noted previously, we will again assume \begin{equation*}
\frac{\chi}{H} \leq 1,
\end{equation*}
Therefore, Boussinesq approximation in the presence of weak  shear implies 
\renewcommand{\theequation}{\arabic{section}.\arabic{equation}}
\begin{align}
\mathrm{either}\,\,\,\,\,\,T_3 \ll T_1 \implies A_t Fr_c \ll Fr,
\label{eq:Bouss_app3}
\end{align}
\begin{align}
\mathrm{or}\,\,\,\,\,\,T_3 \ll T_2 \implies  Fr_c^2 \ll \mathcal{O}(1).
\label{eq:Bouss_app4}
\end{align}
 Since $Fr \ll Fr_c$, that is, the shear is negligible, the condition  $A_t Fr_c \ll Fr$ will require $A_t \ll \mathcal{O}(1)$. Therefore, when shear is  weak, we will  have 
 \begin{equation}
\mathrm{either}\,\,\, A_t \ll \mathcal{O}\left(1\right)\,\, \mathrm{or}\,\,   Fr_c^2   \ll  \mathcal{O}(1)
\label{eq:wave_condition}
\end{equation}
as the  valid conditions for the Boussinesq approximation.   The second clause of   Eq.\ (\ref{eq:wave_condition}) reveals a remarkable fact -- {the Boussinesq approximation may be valid even when $A_t= \mathcal{O}(1)$, provided the background shear is weak enough. Situation like this can arise at the air-sea interface. Hence our finding Eq.\ (\ref{eq:wave_condition})  has broadened  the classical definition of ``Boussinesqness'', conventionally given by Eq.\ (\ref{eq:Bouss_app}) or the first clause of Eq.\ (\ref{eq:wave_condition}).}
 
\subsubsection{Are surface gravity waves Boussinesq or non-Boussinesq?}
\label{sec:sgw_new}
In order to rationalize Eq.\ (\ref{eq:wave_condition}), we consider the case of interfacial gravity waves (existing between two fluids of different densities $\rho_1$ and $\rho_2$) in the absence of background shear, making $Fr\ll Fr_c$ a valid approximation (note that the scaling analysis  will involve $H$, which is now the fluid depth, and not the shear thickness since shear is absent here).  
The phase speed of an intermediate depth Boussinesq interfacial gravity wave is
\begin{equation}
c_B=\pm \sqrt{\frac{A_tg}{k}\Big[1-\exp(-2kH)\Big]},
\label{eq:cb11}
\end{equation}
while the same \emph{without} making the Boussinesq approximation is
\begin{equation}
c_{NB}=\pm \sqrt{ \frac{A_tg}{k}\Big[\frac{1-\exp(-2kH)}{1+A_t \exp(-2kH)}\Big]}.
\label{eq:cb12}
\end{equation}
{The above expressions Eq.\ (\ref{eq:cb11}) and Eq.\ (\ref{eq:cb12}) in a different structure are given in \citet{suth2010} (Eq.\ (2.93) and Eq.\ (2.92) of the book)}. Here $k$ denotes  wavenumber, $A_t\equiv (\rho_2-\rho_1)/(\rho_2+\rho_1)$ is the Atwood number, where $\rho_1$ and $\rho_2$ are respectively the densities of the lighter and the heavier fluids, and $H$ denotes the depth of the heavier fluid (on top of the heavier fluid rests the infinitely deep lighter fluid). It is straight-forward to verify that Eq.\ (\ref{eq:cb11}) is the Boussinesq version of Eq.\ (\ref{eq:cb12}), the former can be obtained by simply substituting $A_t \rightarrow 0$ (when not multiplied with $g$) in the latter.  Taking the ratio of the above two phase speeds, we obtain
\begin{equation}
\frac{c_{NB}}{c_{B}}= \frac{1}{\sqrt{1+A_t \exp(-2kH)}}.
\label{eq:cb13}
\end{equation}
Here we observe that if $A_t \rightarrow 0$, then $c_{NB} \rightarrow c_B$, i.e.\ the Boussinesq approximation is valid, as expected (first clause of Eq.\ (\ref{eq:wave_condition})). However, the Boussinesq approximation is \emph{still} valid if $kH \rightarrow \infty$ (the deep water limit), independent of the value of $A_t$. Hence deep water waves are Boussinesq even when $A_t=1$, i.e.\ \emph{Boussinesq  approximation is valid for  deep water surface gravity waves existing at the air-water interface}. 

Indeed,  the second clause of Eq.\ (\ref{eq:wave_condition}) confirms the observations made from Eq.\ (\ref{eq:cb13}). To see this,  we first write the second clause of Eq.\ (\ref{eq:cb13})
as
$$c\ll \mathcal{O}(\sqrt{gH}),$$
where $c$ denotes the phase speed of the wave, and  substitute $c=c_B$. It shows the Boussinesq approximation to be valid for all wavelengths (long or short), as expected. When $c=c_{NB}$, the Boussinesq approximation is found to be valid in the deep water limit (short waves), \emph{irrespective of $A_t$}. We note here that for shallow water surface gravity waves, Eq.\ (\ref{eq:cb12}) yields $c=c_{NB}=\pm\sqrt{gH}$, implying that the Boussinesq approximation is no longer valid. Equation (2.14) again validates this claim; for shallow water we have $kH \ll 1$, implying
\begin{equation*}
\frac{c_{NB}}{c_{B}} \approx \frac{1}{\sqrt{1+A_t }}+\mathcal{O}(kH),
\end{equation*}
that is, $c_{NB} \nrightarrow c_B$ due to the  explicit dependence on $A_t$.}

We add here that Prof.\ Eyal Heifetz (Tel-Aviv University) has also obtained a similar conclusion by treating surface waves as  vortex sheets and using { the method of image vortices} to take the presence of bottom boundary into account for shallow water surface waves (personal communication).

\subsection{Non-Boussinesq vorticity gravity waves}
\label{sec:NB}
Let us consider a material interface $z=\eta(x,\,t)$ of infinitesimal amplitude, which separates two fluids of different base state properties. The interface  satisfies  kinematic boundary condition, which under linearized assumption yields $\bar{D}\eta/\bar{D}t=w$. This relation, along with the linearized version of Eq.\ (\ref{eq:gov_NSC_1}) produces the following relation  for perturbation density: $\rho=-\eta d\bar{\rho}/dz$. Using these relations, the linearized perturbation vorticity equation Eq.\ (\ref{eq:2.3}) for an interface becomes \citep{heifetz2015stratified}
\begin{align}
\frac{\bar{D}q}{\bar{D}t}=-w\frac{d\bar{q}}{dz}-\frac{g}{\bar{\rho}}\frac{d\bar{\rho}}{dz}\frac{\partial \eta}{\partial x}+\frac{1}{\bar{\rho}^2}\frac{d\bar{\rho}}{dz}\frac{\partial p}{\partial x}.
\label{eq:pert_vor_int_1}
\end{align}

\begin{figure}
\centering\includegraphics[width=100mm]{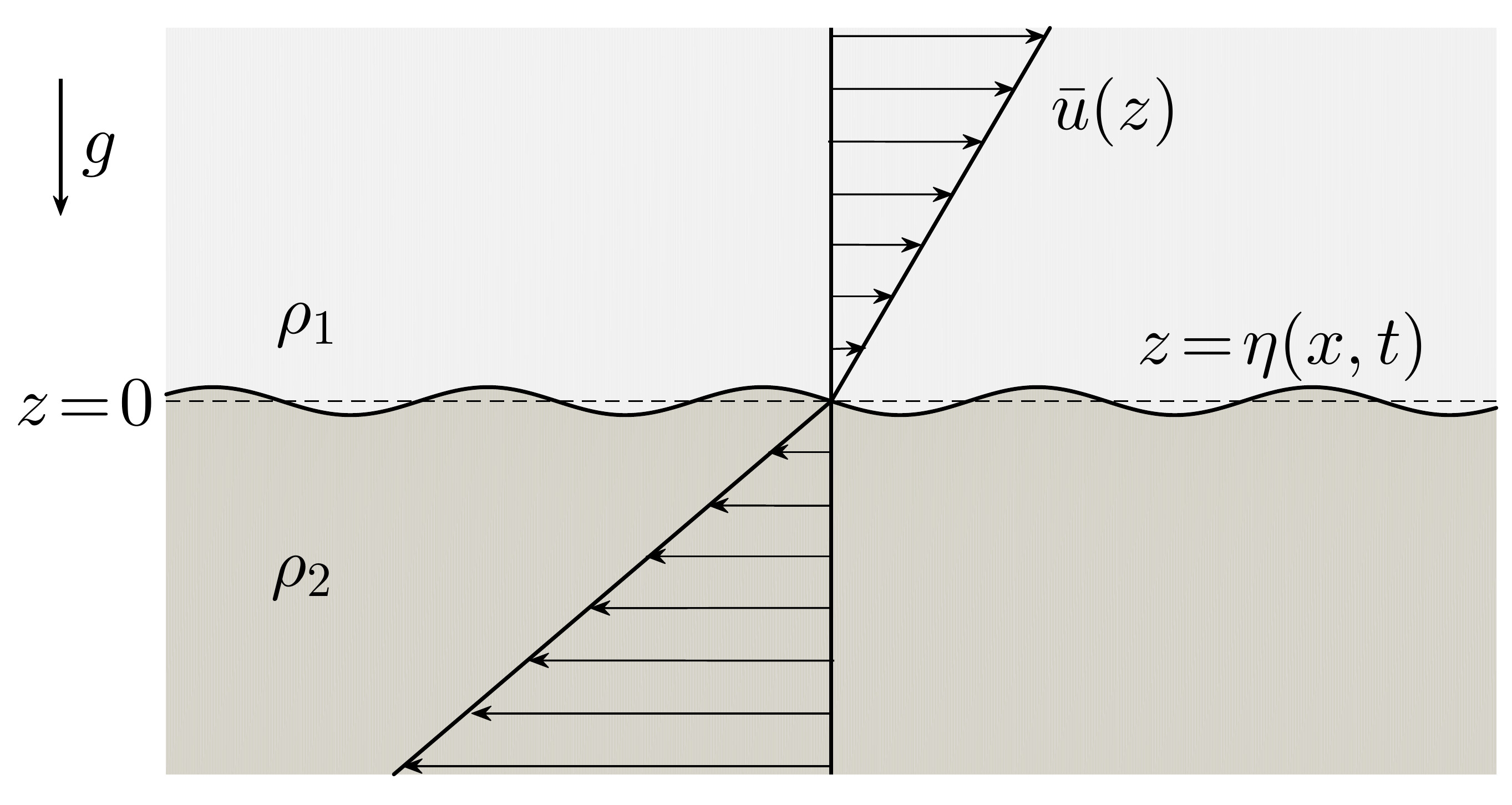}
  \caption{An interface between two fluids of different background density and vorticity in the presence of gravity.}
  \label{fig:1}
\end{figure}

Next we consider a specific case of layerwise constant base state density and vorticity as shown in Fig.\ \ref{fig:1}:
\begin{equation}
\bar{\rho}(z) = \left\{
        \begin{array}{cc}
        \rho_1 & \quad 0 < z \\
        \rho_2 & \quad  z < 0
        \end{array}
    \right.
   \qquad\qquad \bar{q}(z) = \left\{
        \begin{array}{cc}
        Q_1& \quad 0 < z \\
        Q_2 & \quad  z<0.
        \end{array}
    \right.
\end{equation}
 This implies that both $d\bar{q}/dz$ and $d\bar{\rho}/dz$ are delta functions -- zero everywhere except at $z=0$.
Therefore, from \eqref{eq:pert_vor_int_1} we deduce that perturbation vorticity generation takes place only at the interface and the perturbed flow in the bulk is still irrotational (provided, the initial disturbances are also irrotational, which is the case here). Hence we can introduce velocity potentials $(\phi_1,\phi_2)$ separately in the two regions and write the continuity equation
\begin{equation}
\nabla^2\phi_1=0\hspace{0.5cm}z>0\hspace{1cm};\hspace{1cm} \nabla^2\phi_2=0\hspace{0.5cm}z<0\label{eq:Lap1}.
\end{equation}



The linearized kinematic boundary conditions just above and  below the interface, {taking $U=0$ at the interface}, can be written as
\begin{equation}\label{eq:KBC}
\frac{\partial\eta}{\partial t}=\left.\frac{\partial \phi_1}{\partial z}\right|_{z=0}\qquad;\qquad
\frac{\partial\eta}{\partial t}=\left.\frac{\partial \phi_2}{\partial z}\right|_{z=0}.
\end{equation}
The linearized dynamic boundary condition in the presence of piecewise background shear (see Appendix \ref{app:A} for derivation) yields
\begin{equation}\label{eq:DBC}
\rho_1 \left[ \frac{\partial \phi_1}{\partial t}-Q_1\psi+g\eta\right]_{z=0}=\rho_2 \left[ \frac{\partial \phi_2}{\partial t}-Q_2\psi+g\eta\right]_{z=0},
\end{equation}
where $\psi$ is the perturbation streamfunction such that $u=\partial\psi/\partial z$ and $w=-\partial\psi/\partial x$.
A  similar but special case of the above equation, which considered the effect of a constant shear for a free surface gravity wave (air considered as a passive fluid of zero density) in a steady frame, was derived by   \citet{fenton1973some} and subsequently used by  \citet{kishida1988stokes}.
 Using normal mode perturbations for $\phi_1$ and $\phi_2$ in Eq.\ (\ref{eq:Lap1}), and applying the evanescent condition along with the kinematic boundary condition, we get
\begin{equation}
\phi_1=\Re\{A\exp{(-kz)}\exp\left(\ii (kx-\omega t)\right)\}\,\,;\,\,\phi_2=\Re\{-A\exp{(kz)}\exp\left(\ii (kx-\omega t)\right)\},
\end{equation}
where $\omega$ is the complex frequency and $k$ is the wavenumber. The streamfunctions ($\psi_1,\psi_2$) can be recovered using the respective velocity potentials ($\phi_1,\phi_2$).
Substituting the normal modes for velocity potentials, stream functions and surface elevation in Eq.\ (\ref{eq:KBC}) and using Eq.\ (\ref{eq:DBC}), we obtain the dispersion relation for a non-Boussinesq vorticity gravity wave:
\begin{align}
\omega &=\frac{\rho_1 Q_1-\rho_2 Q_2}{2(\rho_1+\rho_2)}\pm\sqrt{ \bigg\{\frac{\rho_2Q_2-\rho_1Q_1}{2(\rho_2+\rho_1)}\bigg\}^2+gk\left(\frac{\rho_2-\rho_1}{\rho_2+\rho_1}\right)}.
\label{eq:non_bouss_vort_grav}
\end{align}
A very similar form of the above equation, which included self-gravity term as well, was obtained in the cylindrical coordinate system for proto-stellar accretion disks  by \citet{yellin2016mechanism}. In the absence of gravity, the above equation will produce a ``vorticity--density wave'' (which becomes a pure vorticity wave when $\rho_1=\rho_2$):
\begin{equation}
\omega=\frac{\rho_1 Q_1-\rho_2 Q_2}{\rho_1+\rho_2}.
\label{eq:vort_waves}
\end{equation}
 Equation (\ref{eq:non_bouss_vort_grav}), under the Boussinesq approximation, produces the dispersion relation for vorticity gravity waves (also given in \citet{harnik2008}):
\begin{equation}
\omega=-\frac{\Delta Q}{4}\pm\sqrt{\left(\frac{\Delta Q}{4}\right)^2+A_tgk},
\label{eq:bouss_vort_grav}
\end{equation}
 where $\Delta Q \equiv Q_2-Q_1$ is the jump in background shear and $A_t\equiv (\rho_2-\rho_1)/(\rho_2+\rho_1)$ is the Atwood number. Pure vorticity wave can be obtained if $A_tg=0$, and pure interfacial gravity waves propagating in opposite directions can be obtained if $\Delta Q=0$. The special case of $A_t=1$ gives the well known dispersion relation for deep water surface gravity waves. This fact  echoes the inference drawn in Sec. \ref{sec:sgw_new} -- the Boussinesq approximation is applicable for deep water surface waves.
 
Unlike the Boussinesq version Eq.\ (\ref{eq:bouss_vort_grav}), the non-Boussinesq dispersion relation Eq.\ (\ref{eq:non_bouss_vort_grav}) shows that shear too, along with gravity, is affected by the variations in density. This non-Boussinesq effect is prominent at higher values of  Atwood number or at higher shear values. The special case of uniform shear (i.e. $Q_1=Q_2=Q$) yields
 \begin{align}
 \omega=-\frac{A_tQ}{2}\pm\sqrt{\left(\frac{A_tQ}{2}\right)^2+A_tgk}.\label{Disp_Q}
 \end{align}
We refer to such waves as ``shear--gravity waves''. Furthermore, we note here that the non-trivial effect of constant shear is not revealed under Boussinesq approximation since the latter only reflects jump in shear, see Eq.\ (\ref{eq:bouss_vort_grav}).  Moreover, in  situations where the effect of gravity is zero or negligible (e.g.\ flow in the horizontal plane), we will still obtain a neutrally propagating wave satisfying the dispersion relation
\begin{align}
 \omega=-A_tQ.
 \label{eq:dens_wave}
\end{align}
 Hence the conventional notion that an interface can only support interfacial waves when there is a jump in buoyancy and/or vorticity across the interface is incomplete. The above equation clearly demonstrates that constant shear across an interface between two fluids of different densities, even in the absence of gravity, can support traveling waves.
 We  refer to such a wave  ``shear--density wave''. Similar wave has also been obtained by \citet{behzad2014role} (they refer to it as ``density wave'')  as well as \citet{yellin2016mechanism} (they have obtained the cylindrical coordinate version, which is more relevant for accretion disks).
  We intend to understand the generation mechanism of shear--density waves, which is not yet properly understood. To this effect we combine Eq.\ (\ref{eq:gov_NSC_12}) with Eq.\ (\ref{eq:pert_vor_int_1}) to obtain
 \begin{align}
\frac{\bar{D}q}{\bar{D}t}=-\frac{1}{\bar{\rho}}\frac{d(\bar{\rho}\bar{q})}{dz}w-\frac{1}{\bar{\rho }}\frac{d(\bar{\rho} g)}{dz}\frac{\partial \eta}{\partial x}-\frac{1}{\bar{\rho}}\frac{d\bar{\rho}}{d z}\frac{\bar{D}u}{\bar{D} t}.
\label{eq:vort_123}
\end{align} 
In the deep water limit, the third term in the RHS is negligible. The second term in the RHS signifies the ``gravitational baroclinic torque''. Under the Boussinesq approximation the first term on the RHS, which is a combination of both barotropic and non-Boussinesq baroclinic torques, becomes zero if the shear $\bar{q}$ is constant.  
Just like $g$ in the second term, a constant $\bar{q}$ in the first term can still generate vorticity.
Constant $\bar{q}$ implies that the first term is proportional  to $\bar{q}d\bar{\rho}/dz$, which leads to the generation of shear--density wave in Eq.\ (\ref{eq:dens_wave}). 
It is straightforward to conclude that both $\bar{q}$ and $d\bar{\rho}/dz$ have to be non-negligible for this wave to exist.
We also infer that it is the variation in the combination $\bar{\rho}\bar{q}$ (and not merely a variation in $\bar{q}$) that causes the waves to propagate, which is evident both from \eqref{eq:vort_123} and from the dispersion relation \eqref{eq:non_bouss_vort_grav}; variation in $\bar{\rho}\bar{q}$ is a necessary and sufficient condition for  the existence of `non-gravity' waves in  stratified shear flows.

\section{Neutral stratification - Instability of a jet penetrating into a  quiescent fluid of a different density}
\label{sec:tr_jet}

\begin{figure}
\centering\includegraphics[width=80mm]{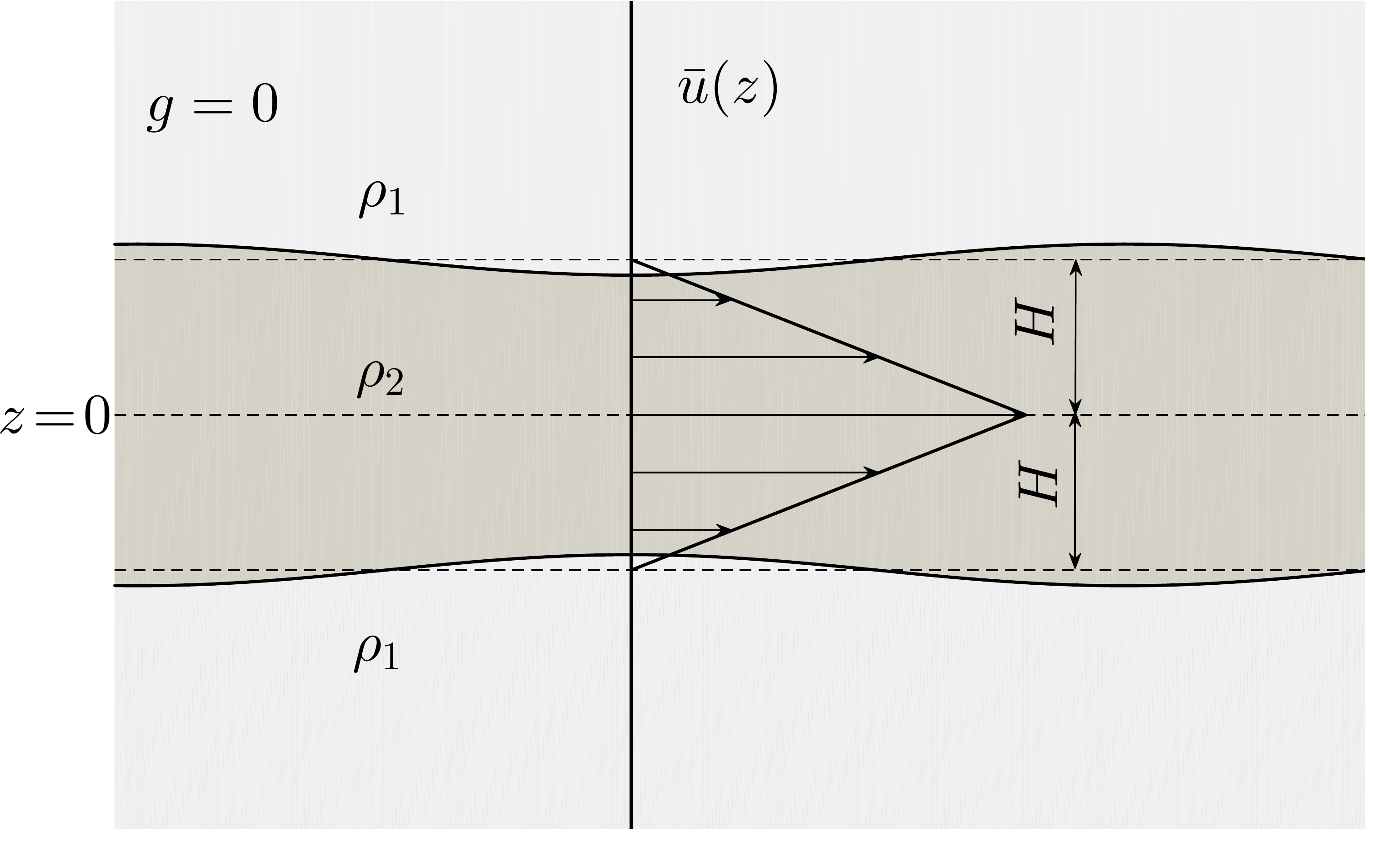}
\caption{Schematic of the triangular jet flow.}
\label{fig:JetS}
\end{figure}

Here {we study} non-Boussinesq instabilities in a particular flow problem of practical relevance. The objective here is to highlight the effect of density stratification in shear instabilities, as opposed to buoyancy stratification, which is commonly studied. The model problem chosen here is that of a triangular fluid jet of density $\rho_2$ penetrating through an ambient fluid (at rest) of density $\rho_1$. The density and velocity profiles are given as
\begin{equation}
\label{eq:triag_jet}
\bar{\rho}(z) = \left\{
        \begin{array}{cc}
        \rho_1 & \quad H < z \\
        \rho_2 & \quad  -H<z < H \\
        \rho_1 & \quad z<-H
        \end{array}
    \right.
   \qquad\qquad \bar{u}(z) = \left\{
        \begin{array}{cc}
        0& \quad H < z \\
        U(1-|z|/H)  & \quad  -H<z< H \\
        0 & \quad z<-H.
        \end{array}
    \right. 
\end{equation}
The flow occurs in the horizontal plane, implying the flow is neutrally stratified, i.e.\ $g=0$. The same situation can be obtained even in the presence of $g$, provided the shear Froude number ($Fr\equiv U/\sqrt{gH}$) is very high (hereafter, we refer to shear Froude number simply as  Froude number).  The system has vorticity--density interfaces at $z=\pm H$ (each interface supporting a  vorticity--density wave) and one vorticity interface at $z=0$ (supporting a vorticity wave); see Fig.\ \ref{fig:JetS}. As is now well established, interfacial waves can resonantly interact amongst themselves leading to normal mode instabilities \citep{carp2012,guha2014wave} if they form a counter-propagating configuration \footnote{In a counter-propagating system of two waves (each present at its own interface), the intrinsic phase speed of the waves should be opposite to each other. Additionally, each wave's intrinsic phase speed should be opposite to the local mean flow (unless the local mean velocity is zero).}. Hence we can expect that these waves - two vorticity--density waves can interact with the counter-propagating vorticity wave to yield normal mode instability.

The unstratified version of Eq.\ (\ref{eq:triag_jet}), given in \citet{drazin2002introduction} and \citet{carpenter2011instability}, reveals that
only the {wavenumbers $(kH)$} less than $1.83$ are unstable.
Using the procedure  outlined in Sec. \ref{sec:gov_eq}, we obtain the following dispersion relation in terms of non-dimensional frequency $\tilde{\omega}\equiv \omega/(U/H)$ and wavenumber $\alpha \equiv kH$ : 
\begin{equation}
\left(\tilde{\omega}+\frac{1}{R+\tanh{\alpha}}\right)\left(\tilde{\omega}^2+\frac{\alpha R+\alpha\tanh{\alpha}-R\tanh{\alpha}}{R+\tanh{\alpha}}\tilde{\omega}+\frac{\alpha-\tanh{\alpha}}{R+\tanh{\alpha}}\right)=0,
\end{equation}
where  $R\equiv\rho_1/\rho_2$ is the density ratio, and is related to the Atwood number by 
$$R=\frac{1-A_t}{1+A_t}.$$
It can be seen that one of the three roots is always real while the other two can be either real or complex conjugates depending on the values of $R$ and $\alpha$. The region of instability is given by
\begin{equation}
R<\frac{-\alpha\tanh{\alpha}+2+2\sech{\alpha} }{\alpha-\tanh{\alpha}}.
\end{equation}
\begin{figure}
\centering\includegraphics[width=80mm]{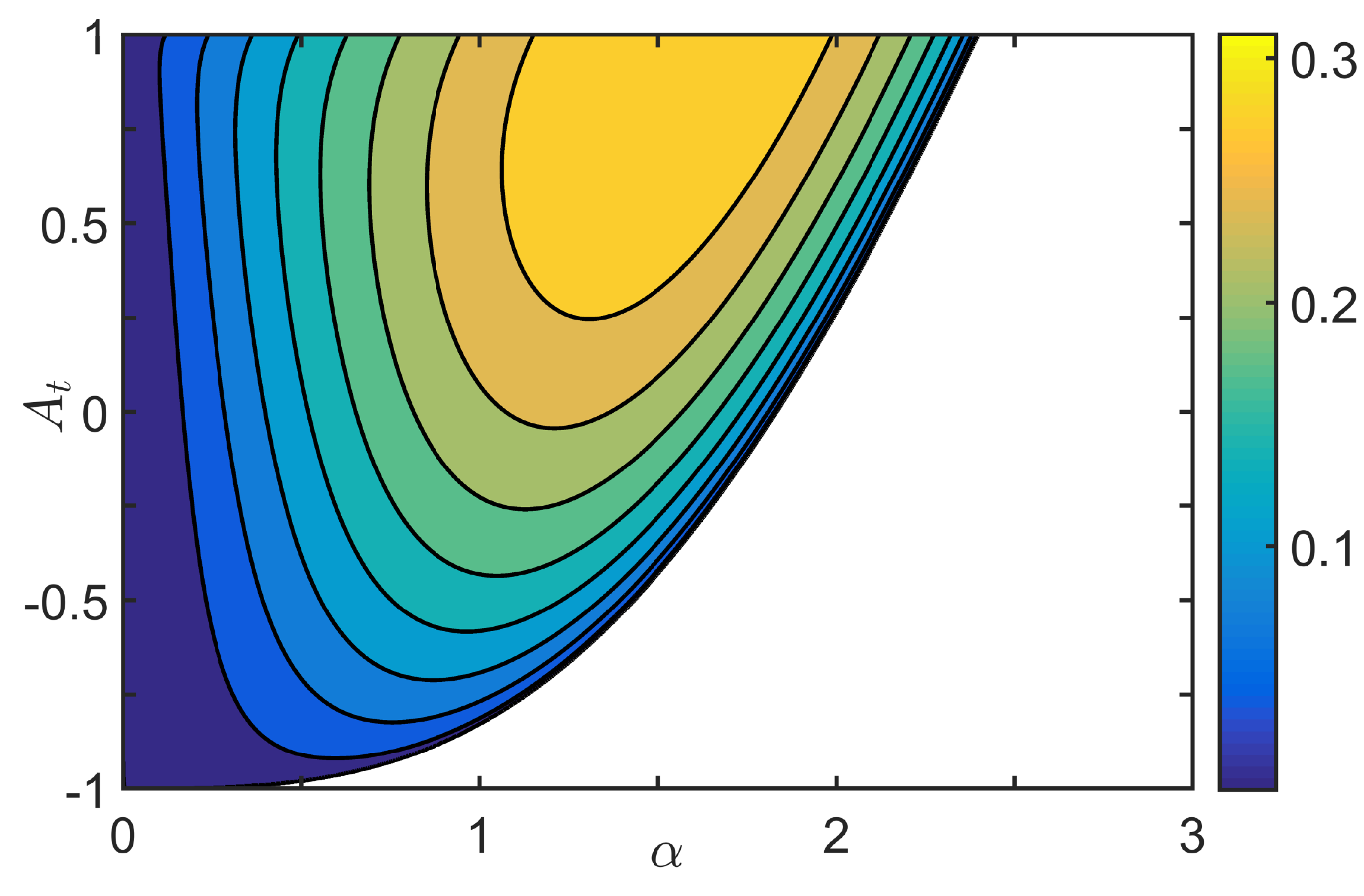}
\caption{$A_t-\alpha$ plot showing growth rate contours for the triangular jet flow.}
\label{fig:Jet}
\end{figure}

\noindent When $A_t=0$ (homogeneous case), the maximum growth rate is $\gamma_{max}=0.2470$  (occurs at $\alpha_{max}=1.2258$) and the cutoff wavenumber is $\alpha_{cut}\approx 1.833$, both confirming the results in \citet{drazin2002introduction}.
Fig.\ \ref{fig:Jet} shows that as the density of the middle layer increases with respect to the surrounding, the growth rate as well as the cutoff wavenumber also increases. For $A_t\rightarrow 1$, maximum growth rate of $0.3098$ is obtained at $\alpha_{max}=1.6063$, which is significantly higher than the homogeneous growth rate. This is also the maximum growth rate in {the $A_t-\alpha$ parameter space.} The cut-off wavenumber for $A_t\rightarrow 1$ is $\alpha_{cut}=2.399$, whereas for $A_t\rightarrow-1$, the cutoff wavenumber approaches zero.

 
 \section{Instabilities of a shear layer with stably/neutrally stratified density interface }
 \label{sec:KH}
The objective of this section is to understand shear instabilities occurring in non-Boussinesq shear layers. We consider the classic Holmboe configuration (Fig.\ \ref{fig:KH}(a)) with and without gravity. As previously mentioned, the situation of zero gravity can also be interpreted as a situation with a very high Froude number.  For the case of non-zero gravity, we will  restrict ourselves to the case of a stable density stratification.  We note that when $g=0$, the familiar concepts of static stability (lighter fluid on the top of heavier fluid) and static instability (heavier fluid on the top of lighter fluid) make little sense. The effect of density stratification is kinematic instead of dynamic -- on flipping the density stratification about $z=0$, the interfacial wave travels in the opposite direction.
Hence, without a loss of generality we will assume $A_t>0$. Firstly, we will study the case of heterogeneous Kelvin-Helmholtz instability (KHI) in zero gravity, and then non-Boussinesq Holmboe instability in {the} presence of gravity. 
 
 \subsection{Neutral stratification - Kelvin-Helmholtz instability with a density interface}


\begin{figure}
        \centering
        		\begin{subfigure}[b]{0.48\textwidth}
                \centering
                \includegraphics[width=\textwidth]{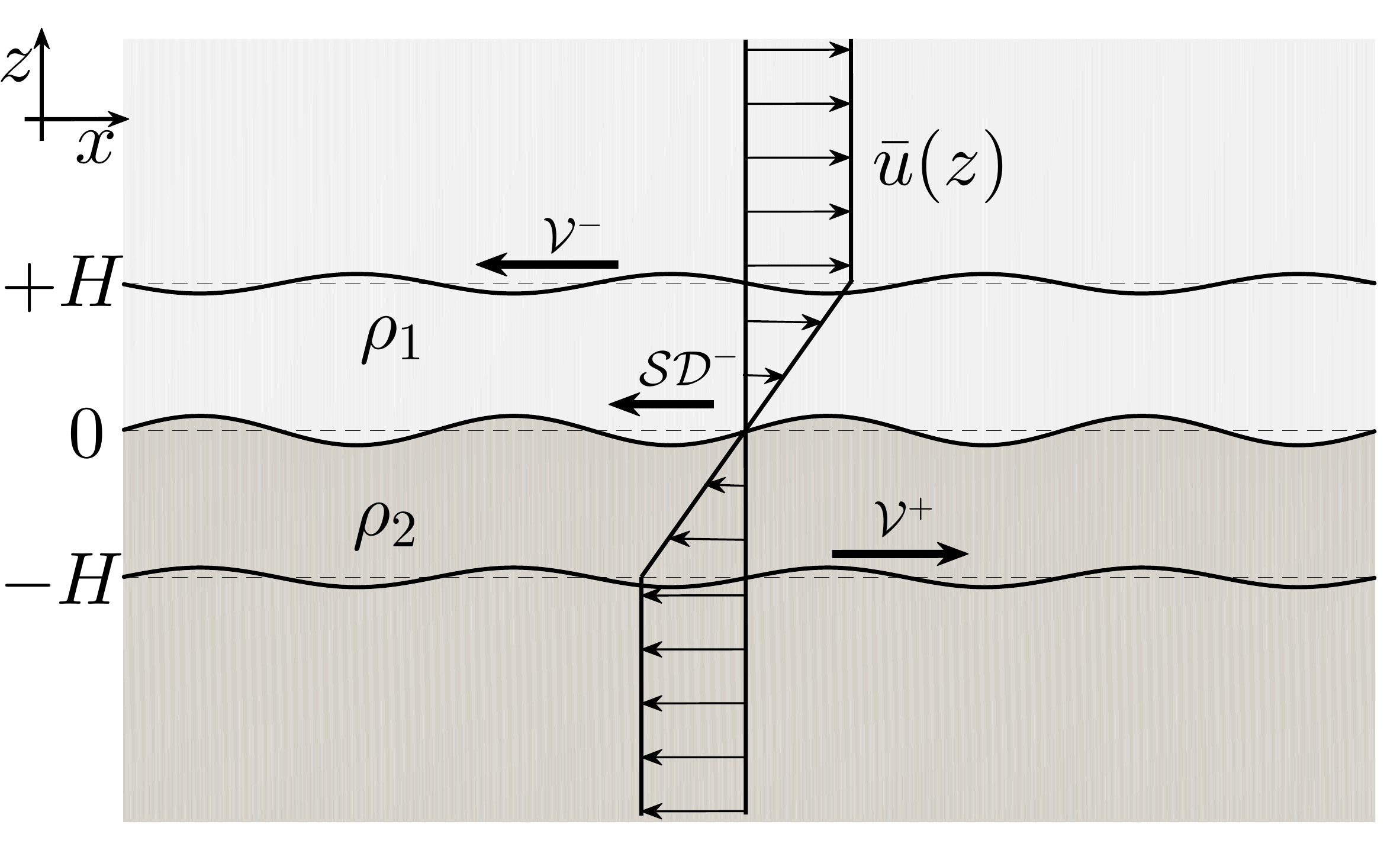}
                \label{fig:KH2}
        \end{subfigure}%
        \begin{subfigure}[b]{0.48\textwidth}
                \centering
                \includegraphics[width=\textwidth]{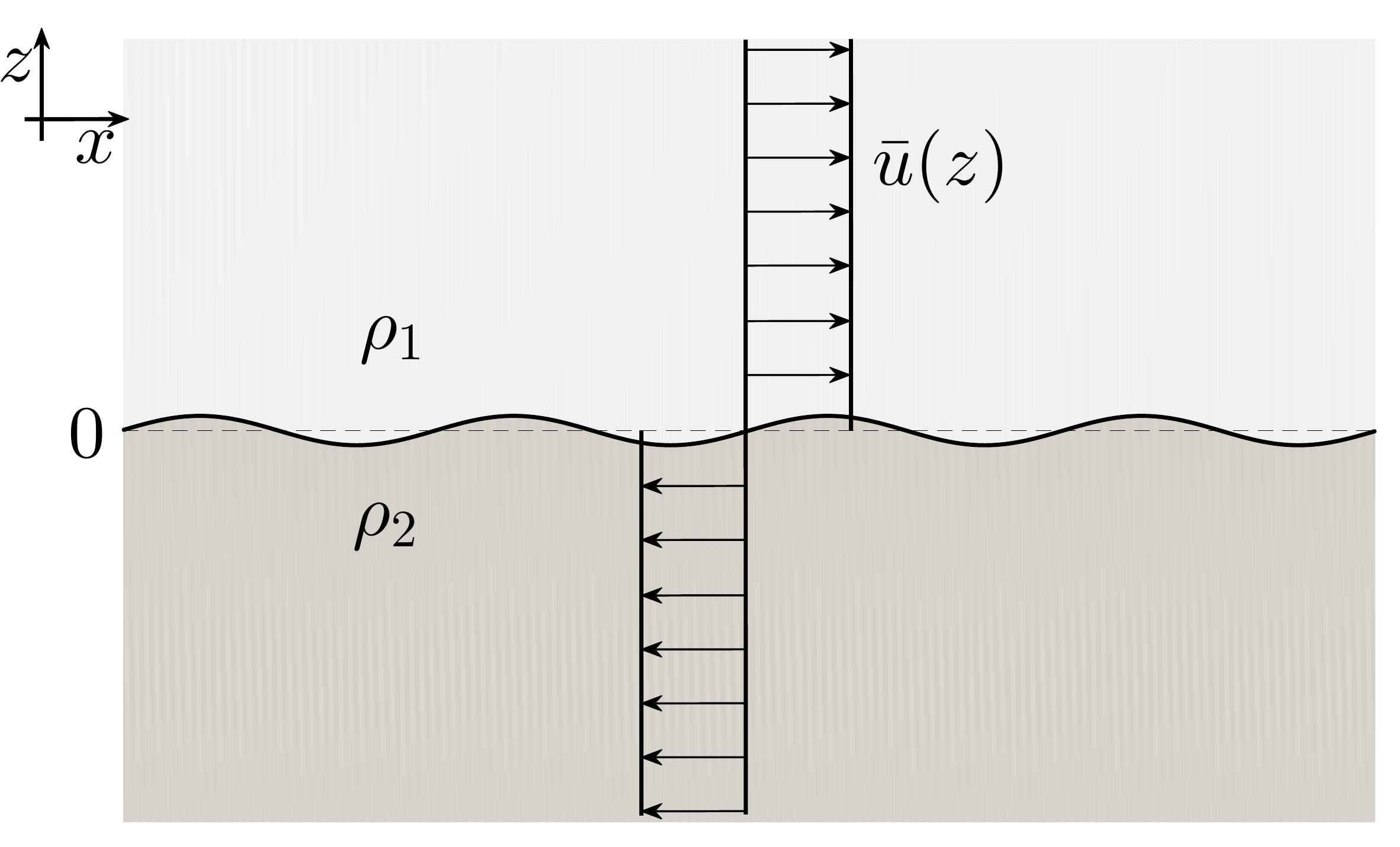}
                \label{fig:KH1}
        \end{subfigure}%
                \caption{(a)  Shear layer of finite thickness with a density interface. (b) Shear layer of infinitesimal thickness -- the limiting condition of (a).}

        \label{fig:KH}
\end{figure}

A homogeneous Kelvin-Helmholtz instability (KHI) or an instability of an unbounded vortex sheet \citep{drazin2002introduction}, in the absence of gravity, is a well studied phenomenon in fluid dynamics. This instability is understood to be a limiting case of instability of an unbounded shear layer (Fig.\ \ref{fig:KH}(a)), when the thickness of shear layer tends to zero (Fig.\ \ref{fig:KH}(b)). Although the KHI set-up in Fig.\ \ref{fig:KH}(a) is of mathematical relevance in itself, it can also be observed in absence of gravity in various types of flows, e.g.\ \citet{ofman2011sdo,behzad2014role}.   In terms of resonant wave interaction, this instability (both for homogeneous and Boussinesq cases) is explained in terms of two mutually interacting vorticity waves \citep{carpenter2011instability}. The instability of the  heterogeneous unbounded vortex sheet in the absence of gravity (or for very high shear) has also been studied (Fig.\ \ref{fig:KH}(b)), the dispersion  relation of which can be written as \citep{drazin2002introduction}:


\begin{equation}
\omega=-kA_tU\pm\ii \sqrt{(1-A_t^2)(k U)^2},\label{eq:KH1}
\end{equation}
where the top (bottom) layer velocity is $U$ ($-U$). 
{It is evident from \eqref{eq:KH1} that} as the density difference (characterized by $A_t\equiv (\rho_2-\rho_1)/(\rho_2+\rho_1)$) increases between the two layers, the growth rate $\sqrt{(1-A_t^2)(Uk)^2}$ diminishes and the flow becomes stabilized. In the case of a homogeneous flow, there exists only two vorticity waves; however, in the case of a heterogeneous flow (two-layered flow to be specific),  the system undergoes substantial change due to the presence of a shear--density wave on the density interface, which may significantly alter the dynamics of the system. The dispersion relation Eq.\ (\ref{eq:dens_wave}) shows that if both $A_t$ and  shear are positive, we will get a negatively propagating shear density wave, represented by $\mathcal{SD^-}$ in Fig.\ \ref{fig:KH}(a). We note here that if $A_t<0$, then the wave will be $\mathcal{SD^+}$ instead of $\mathcal{SD^-}$, which {does not} change the dynamics of the problem {in the absence of gravity}. A necessary criteria for instability based on wave interaction theory is that the involved neutral waves, when taken separately, should be propagating in the opposite directions relative to each other. From  Fig.\  \ref{fig:KH}(a), it is evident that there can be two possible pairs of such interactions - (i)  between the vorticity waves $\mathcal{V^+}$ and $\mathcal{V^-}$ {(i.e. the Rayleigh mode instability)}, and (ii)  between the vorticity wave $\mathcal{V^+}$ and the shear--density wave $\mathcal{SD^-}$. The other necessary criteria for instability i.e., the local mean flow is opposite to the direction of the phase speed for each wave, is also  satisfied by both the pairs. Therefore for the heterogeneous case, we will obtain two different types of interactions. This is quite different from the homogeneous case, in which only the Rayleigh mode instability is obtained. The dispersion relation for the heterogeneous instability is given by
\begin{align}\label{eq:KH2}
\tilde{\omega}^3+C_2\tilde{\omega}^2+C_1\tilde{\omega}+C_0=0.
\end{align}
Here $\;\tilde{\omega}\equiv\omega/(U/H)$ and
$$C_2=2A_t\tanh{\alpha}(\tanh{\alpha}+1)^{-1};\quad C_1=[\alpha-\alpha^2-\tanh{\alpha}(\tanh{\alpha}+1)^{-2}];$$ $$C_0=-A_t[\alpha-\tanh{\alpha}(\tanh{\alpha}+1)^{-1}]^2;\,\,\,\,\,\,\,\,\,\,\,\,\,\,\,\,\,\,\,\,\,\,\,\,\,\,\,\,\,\,\,\,\,\,\,\,\,\,\,\,\,\,\,\,\,\,\,\,\,\,\,\,\,\,\,\,\,\,\,\,\,\,\,\,\,\,\,\,\,\,\,\,\,\,\,\,\,\,\,\,\,\,\,\,\,\,\,\,\,$$ where $\alpha\equiv kH.$ 
\begin{figure}
        \centering
        \begin{subfigure}[b]{0.45\textwidth}
                \centering
                \includegraphics[width=\textwidth]{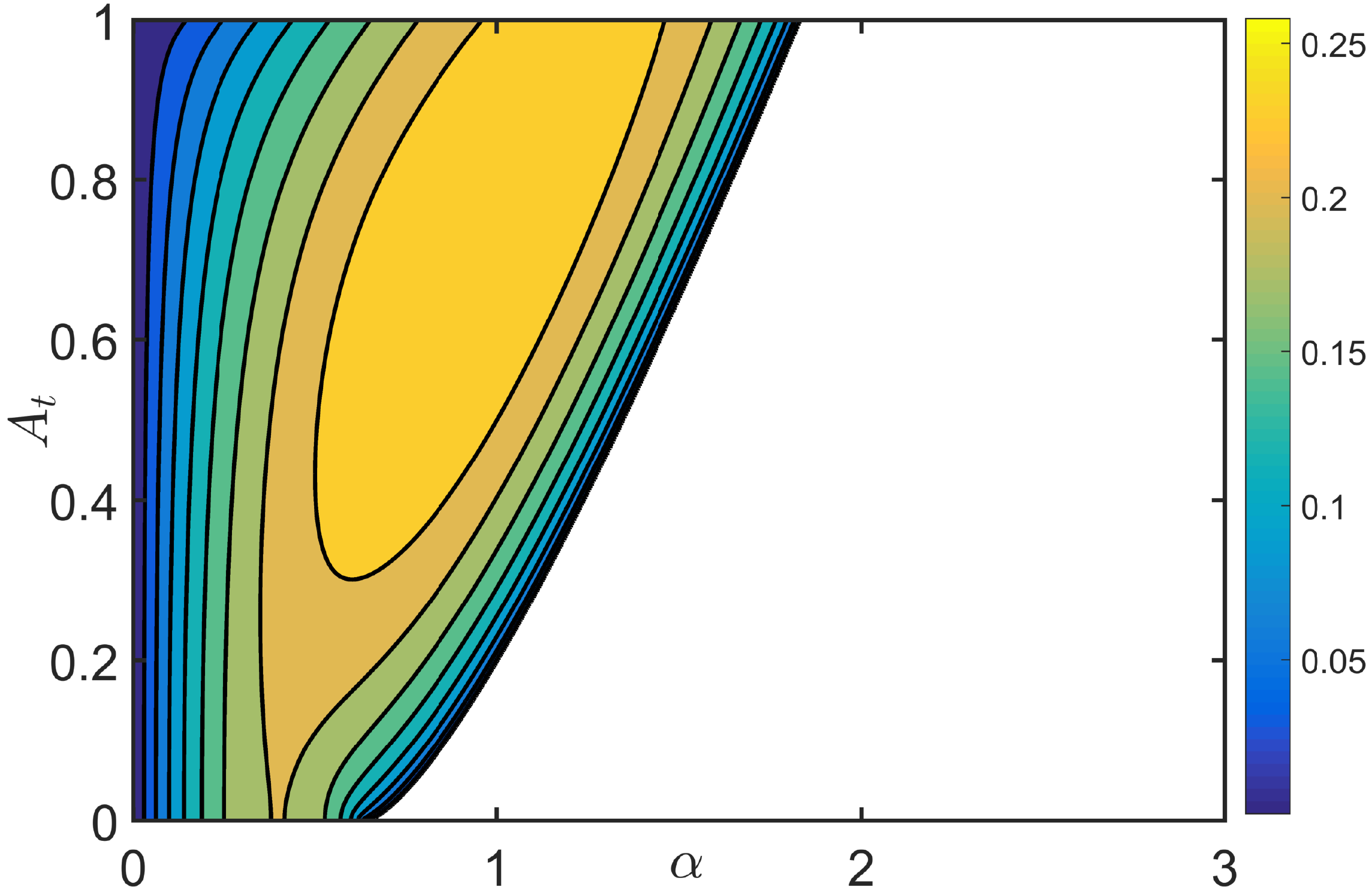}
                \caption{}
        \end{subfigure}%
		\begin{subfigure}[b]{0.45\textwidth}
                \centering
                \includegraphics[width=\textwidth]{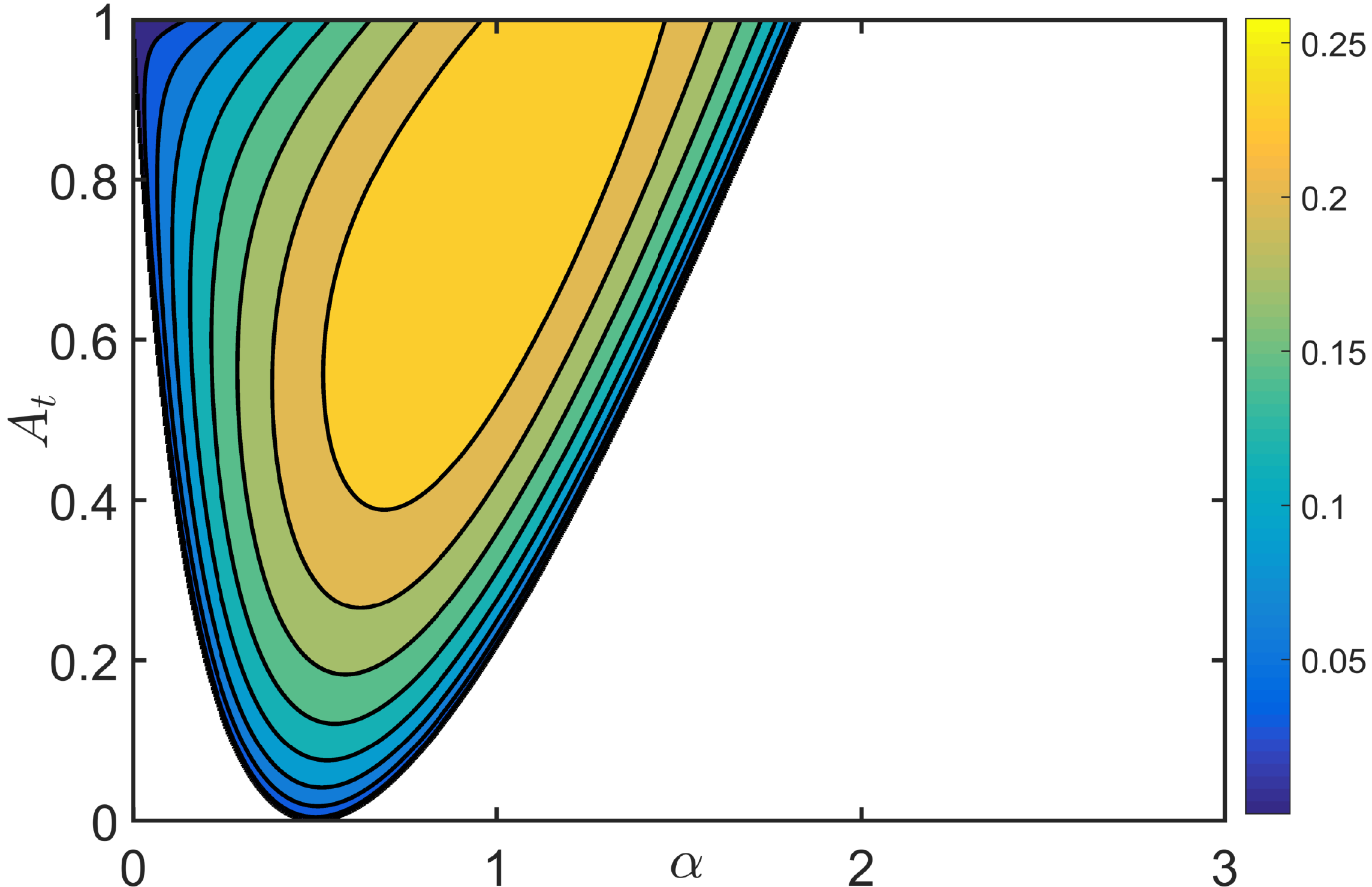}
                \caption{}
       \end{subfigure}%
        \caption{$A_t-\alpha$ plot showing growth rate contours when (a)  all three waves are present in the system, and  (b) wave $\mathcal{V^-}$ (i.e.\ interface at $z=H$) is removed. }
\label{fig:KH33}
\end{figure}
\begin{figure}
        \centering
        \begin{subfigure}[b]{0.45\textwidth}
                \centering
                \includegraphics[width=\textwidth]{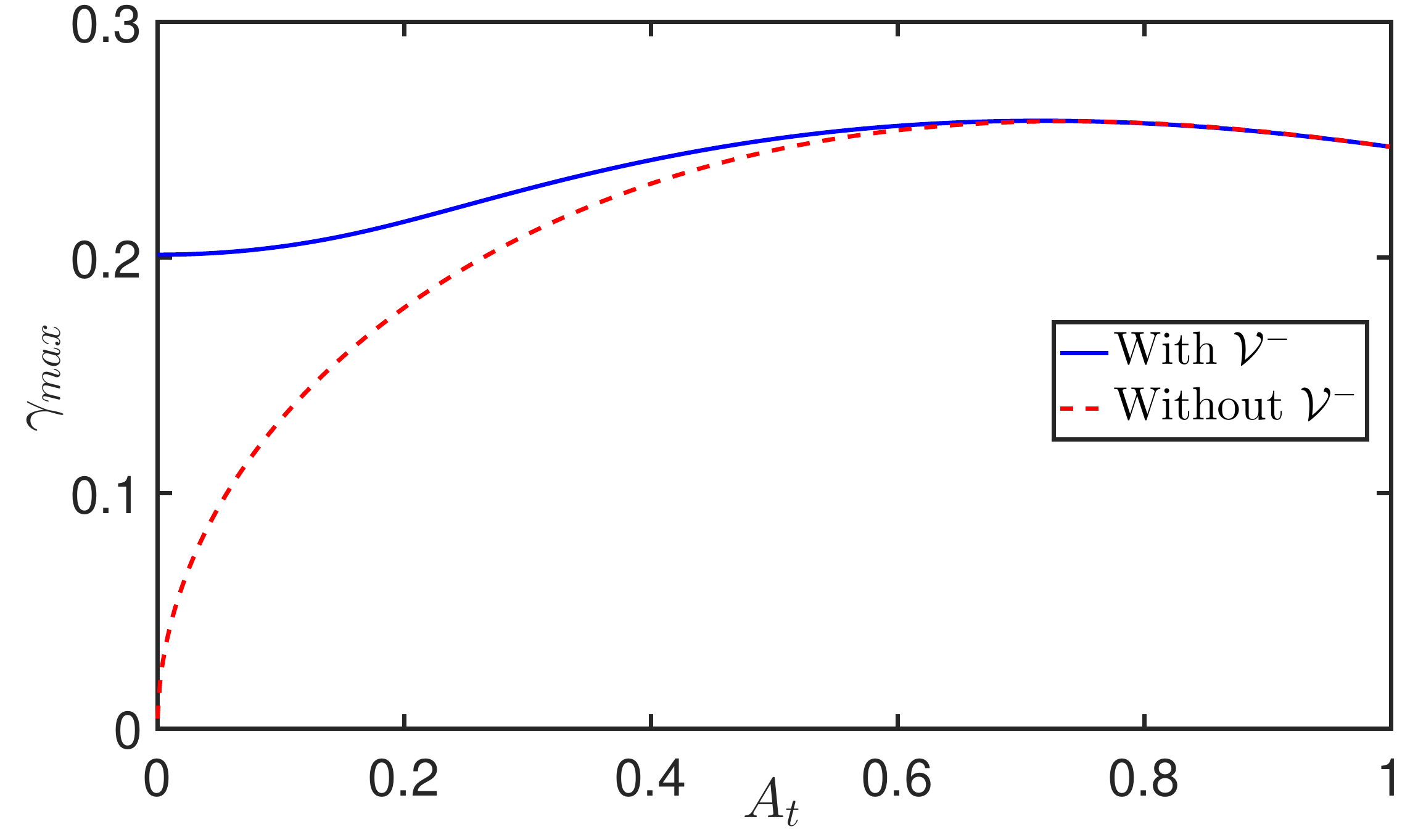}
   \caption{}
        \end{subfigure}%
		\begin{subfigure}[b]{0.45\textwidth}
                \centering
                \includegraphics[width=\textwidth]{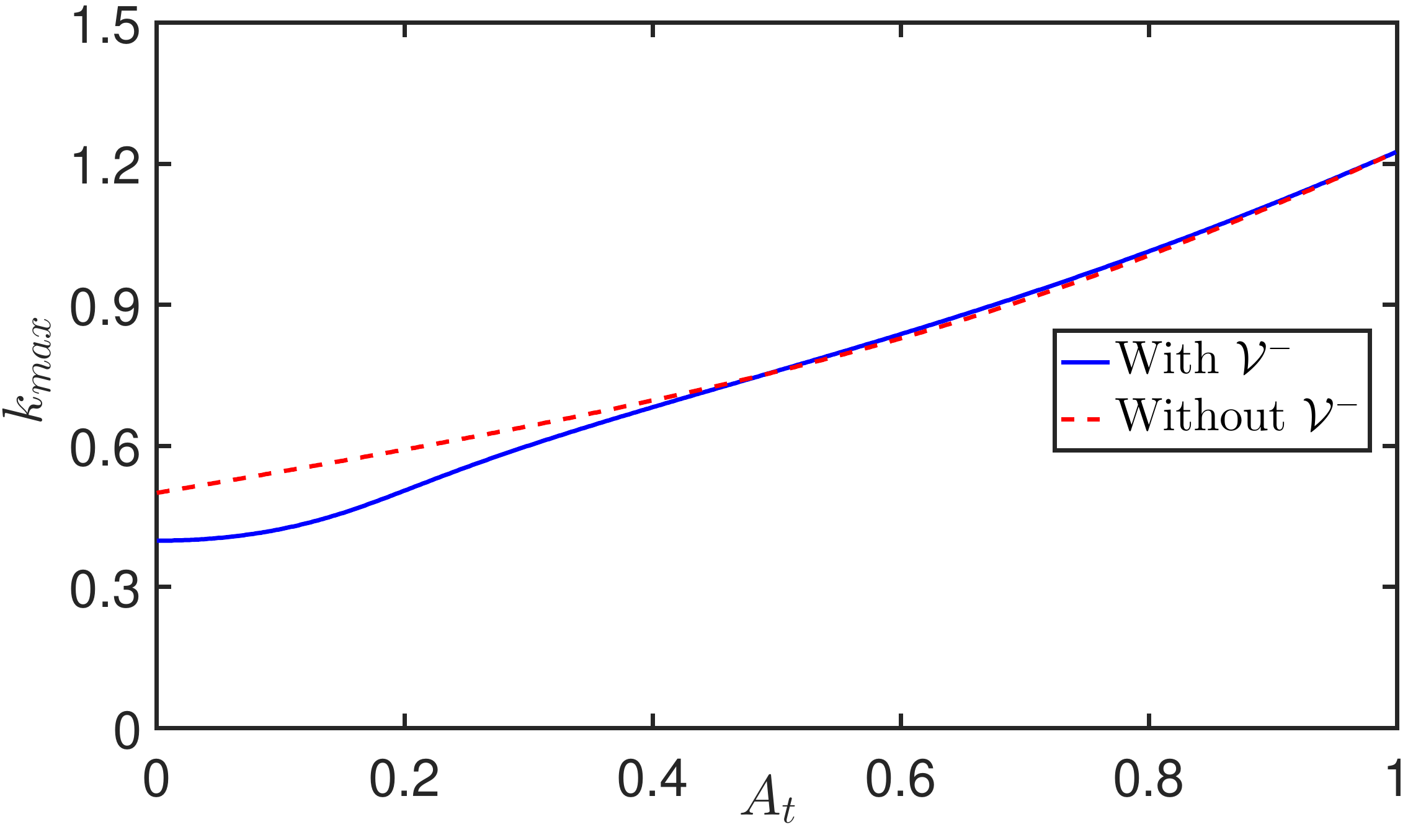}
  \caption{}
        \end{subfigure}%
        \caption{Plot between (a) maximum growth rate ($\gamma_{max}$) and (b) wavenumber of maximum growth $(k_{max})$  versus Atwood number $(A_t)$.}
        \label{fig:KH555}
\end{figure}
Using the dispersion relation, we plot the growth rate contours in $A_t-\alpha$ plane (Fig.\ \ref{fig:KH33}(a)). From the growth rate plot we see that for $A_t=0$, all wavenumbers lesser than 0.64 are unstable, which is expected {for the homogeneous KHI}\citep{drazin2004hydrodynamic}. The  dominant interaction here is known to be between the two vorticity waves. However, as  $A_t$ keeps on increasing, more and more wavenumbers get destabilized. This observation is a hint that for higher $A_t$, the instability might be  dominated by the interaction between $\mathcal{V^+}$ and  $\mathcal{SD^-}$. In order to conclusively study this interaction, we eliminate the wave $\mathcal{V^-}$ by assuming a uniform shear for $z>0$, and then draw the growth rate contours (note we are left with  only two waves -- $\mathcal{SD^-}$ and $\mathcal{V^+}$). The growth rate contours (Fig.\ \ref{fig:KH33}(b)) show that  there is a substantial difference for lower values of $A_t$. However as $A_t$ increases, the differences between the two cases (i.e.\ with and without $\mathcal{V^-}$) become negligible. Furthermore, for every $A_t$, we find the maximum growth rate $\gamma_{max}$ and the corresponding wavenumber $k_{max}$ (for each of these two cases) and respectively plot them in Fig.\ \ref{fig:KH555}(a) and  Fig.\ \ref{fig:KH555}(b). Again, in doing so, we notice that the curves for $\gamma_{max}$ for both these cases nearly overlap for $A_t \gtrsim 0.5$. This analysis conclusively shows that the dominant interaction in KHI with a density interface for lower values of $A_t$ in the absence of gravity is between the two vorticity waves. However for higher values of $A_t$, the instability is mainly governed by resonant interaction between the shear density wave and the vorticity wave existing in the denser fluid. Our conclusions are in {qualitative} agreement with that of  \citet{behzad2014role}. 

Unlike the case of instability of a vortex sheet, where $A_t$ had a stabilizing effect,  the  {behavior of growth rate in an unbounded shear layer no longer remains monotonic. For lower Atwood numbers, where the instability is dominated by an interaction between $\mathcal{V^-}$ and $\mathcal{V^+}$, it increases with increasing $A_t$ whereas for higher $A_t$, $\gamma_{max}$ decreases with $A_t$}. Only in the limiting case of infinitesimally small shear layer thickness (i.e. $H\rightarrow 0$), the dispersion relation \eqref{eq:KH2} approaches that of a vortex sheet \eqref{eq:KH1}, and $A_t$ has a stabilizing role. {It might also be noted that for a finite shear layer the maximum growth rate $\gamma_{max}$ for a particular $A_t$ is the maximum of growth rates over all values of $kH$.} {We would also like to point out that when we are in the limit of $H\rightarrow 0$, we have coalesced the three interfaces into only a single interface. Initially, we had three waves in the system (two vorticity and one shear density)  but in the limit of $H\rightarrow 0$ we will only have two  waves which cannot be called purely a vorticity wave or a shear density wave.}

\subsection{ Stable stratification - the Non--Boussinesq Holmboe instability}
\label{sec:Holmb_in}
 
Holmboe instability is usually understood in terms of interaction between an interfacial (Boussinesq) gravity wave and a vorticity wave, locking in phase and giving rise to exponential growth  \citep{baines1994mechanism, carp2012}. The non--Boussinesq Holmboe instability  was studied by  \citet{umurhan2007holmboe} { and later by \citet{barros2011holmboe}}. The work of \citet{umurhan2007holmboe} is more relevant to this paper since they explained the non-Boussinesq effects by including the baroclinic term $T_3$ (of Eq.\ (\ref{eq:2.3})). It was also concluded that the inclusion of the non-Boussinesq term breaks the symmetry of the Holmboe modes, which are otherwise symmetric in the Boussinesq case. This is because in the Boussinesq case, the effect of a  constant shear at the interface $z=0$ will not be felt by the gravity waves, and both the gravity waves on the buoyancy interface will have the same frequency. However, including the non-Boussinesq term will differentiate between the two `gravity' waves at the interface, as is evident from  \eqref{Disp_Q}. The leftward moving gravity wave will be sped up while the rightward moving one would be slowed down. Furthermore, under the Boussinesq approximation,  $A_t$ and $Fr$ are often  consolidated into a single parameter, the Richardson number: $J\equiv A_t/Fr^2$. This makes sense from our result Eq.\ (\ref{eq:Bouss_app}) -- $A_t$ can be held constant (at a small value) while  $Fr$ can be varied.
Such consolidation is not useful for  non-Boussinesq flows.
For  example, $A_t=0.001$ and $Fr=0.2$ will yield $J=0.025$, but $A_t=0.1$ and $Fr=2$ will also yield the same $J$. The first case falls within the Boussinesq limit, while the second case does not,  basically implying that the use of $J$ as a parameter for a non-Boussinesq flow does not offer any simplification. Hence we will treat $A_t$ and $Fr$ as two distinct parameters of our problem.
\begin{figure}
	\centering\includegraphics[width=0.6\linewidth]{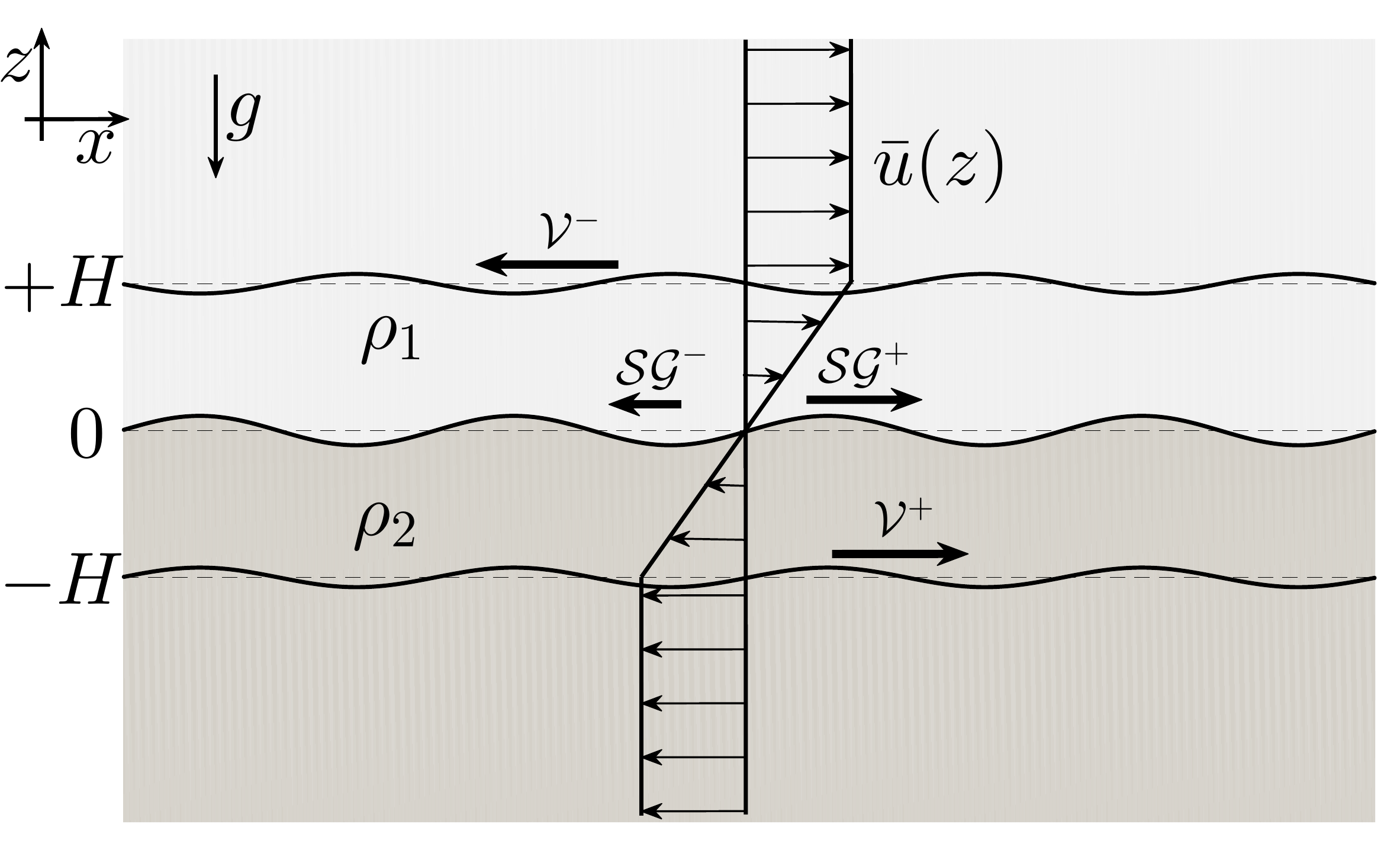}
  	\caption{ Schematic of the Holmboe configuration. { The figure is similar to the Fig. \ref{fig:KH}(a) but with gravity present.}}
  	\label{fig:23}
\end{figure}


In the Holmboe configuration, each interface can support one or more waves (marked in  Fig.\ \ref{fig:23}),  which implies that the waves present at different interfaces can interact among themselves and lead to instability \citep{guha2014wave}. Waves $\mathcal{V^+}$ (rightward) and $\mathcal{V^-}$ (leftward) are the vorticity waves (Rossby edge waves if the reference frame is rotating) and their physics is well known \citep{carpenter2011instability,guha2014wave}. In the Boussinesq limit, waves $\mathcal{SG^-}$ (leftward) and $\mathcal{SG^+}$ (rightward) are simply treated as  gravity waves (here $\mathcal{SG}$ denotes shear--gravity waves), ignoring the effect of the constant shear. However, in  a fully non-Boussinesq setting the two gravity waves can feel the effect of shear {due to density variation across the interface, even though the base shear remains constant across the density interface}. In such a system,  we refer them shear--gravity waves, thereby treating shear at par with gravity. This is because the waves could be sustained independently both by shear as well as by gravity, in presence of a density jump. 
It was shown {by} \citet{lawrence1991stability} for Boussinesq Holmboe instability that Holmboe modes exist for $J>0.07$, in this case there exists 2 pairs of complex conjugate roots - one of which exits due to an interaction between wave $\mathcal{V^-}$ and wave $\mathcal{SG^+}$, while the other one is due to an interaction between wave $\mathcal{V^+}$ and wave $\mathcal{SG^-}$. However, for $J<0.07$, the instability modes are termed as Kelvin-Helmholtz (KH) modes with a reasoning that the instability has no phase speed of propagation (i.e. roots are imaginary). 
Apart from looking at the dispersion relation, the difference between KH mode and Holmboe mode can also be explained by the physical understanding of the Richardson number. Richardson number can be seen as the ratio of buoyant forces and the shear forces in a flow. This means that at lower $J$, the buoyant forces are sub-dominant and the role of vorticity waves in the instability prevails over the role of vorticity waves. Further, the case of $J\rightarrow 0$ is treated as the case of $A_t\rightarrow 0$ and thus, at lower $J$, the flow approaches the classical (unstratified) Rayleigh shear profile. However, we find that in a non-Boussinesq flow, this might not be the case. We expect that when the flow is shear dominated, the flow will behave more like the flow described in Sec. \ref{sec:KH}, which effectively means that also at higher Froude numbers, the instability may not be dominated by an interaction between two vorticity waves $\mathcal{V^+}$ and $\mathcal{V^-}$, despite buoyancy effects being minimal. According to {our analysis}, the underlying reason is as follows -  the shear--gravity wave $\mathcal{SG^-}$ tends to behave like shear--density wave $SD^-$ at higher Froude numbers (which is evident from the dispersion relation Eq.\ (\ref{Disp_Q}), whereas the speed of the other wave $\mathcal{SG}^+$ starts going towards zero, which is evident from the dispersion relation of isolated shear--gravity waves (see Eq.\ (\ref{eq:bouss_vort_grav})). Therefore, at higher Froude numbers and Atwood numbers, even for low buoyant forces, the instability will be dominated by the interaction between waves $\mathcal{SG^-}$ and $\mathcal{V^+}$. 
\begin{figure}
	\centering\includegraphics[width=120mm]{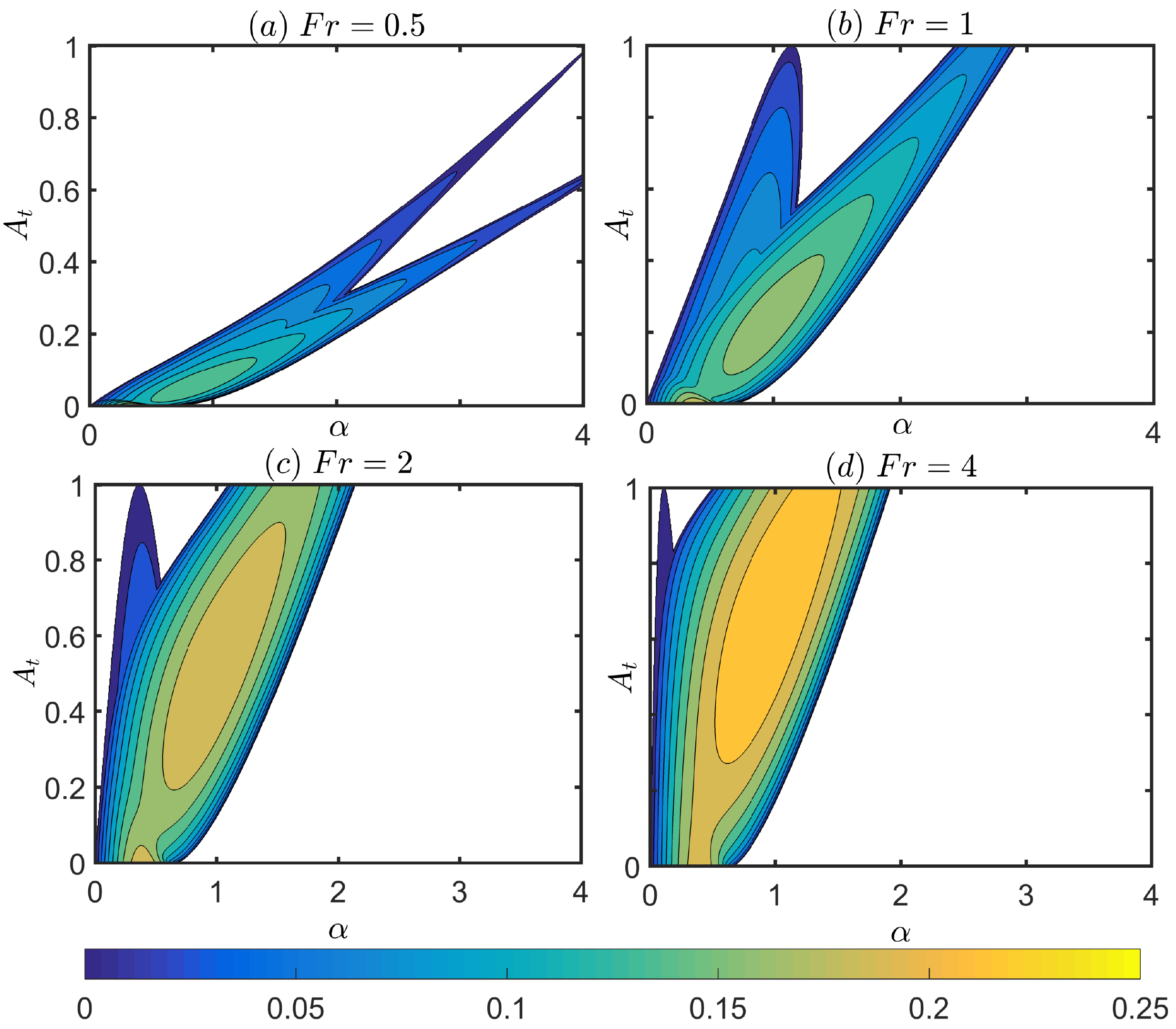}
  	\caption{ $A_t$--$\alpha$  plot showing non-dimensionalised growth rate $(\gamma/(U/H))$ contours for non-Boussinesq Holmboe configuration for different $Fr$.}
  	\label{fig:Holmboe}
\end{figure} 
We plot the growth rate contours for a non-Boussinesq Holmboe instability for different Froude numbers (Fig.\ \ref{fig:Holmboe}). From the figures, we can validate that for the non-Boussinesq Holmboe modes, the symmetry between rightward and leftward propagating Holmboe breaks down \citep{umurhan2007holmboe}. At higher Atwood numbers (especially Fig.\ \ref{fig:Holmboe}(a)), this leads to two distinct regions of instability because the left Holmboe and the right Holmboe gets separated. {As mentioned previously,  the phase speed of $\mathcal{SG^-}$ increases with an increase in the shear} and hence it becomes easier for $\mathcal{SG^-}$  to lock in phase with $\mathcal{V^+}$, while on the other hand, the phase speed of $\mathcal{SG^+}$ goes towards zero making the interaction between $\mathcal{V^-}$ and $\mathcal{SG^+}$ weaker.  Since it's not directly inferable from the Fig.\ \ref{fig:Holmboe}  which region belongs to which interaction, we have again eliminated wave $\mathcal{V^-}$ and $\mathcal{V^+}$ respectively, one at a time.

\begin{figure}
        \centering
        \begin{subfigure}[b]{\textwidth}
                \centering
                \includegraphics[width=100mm]{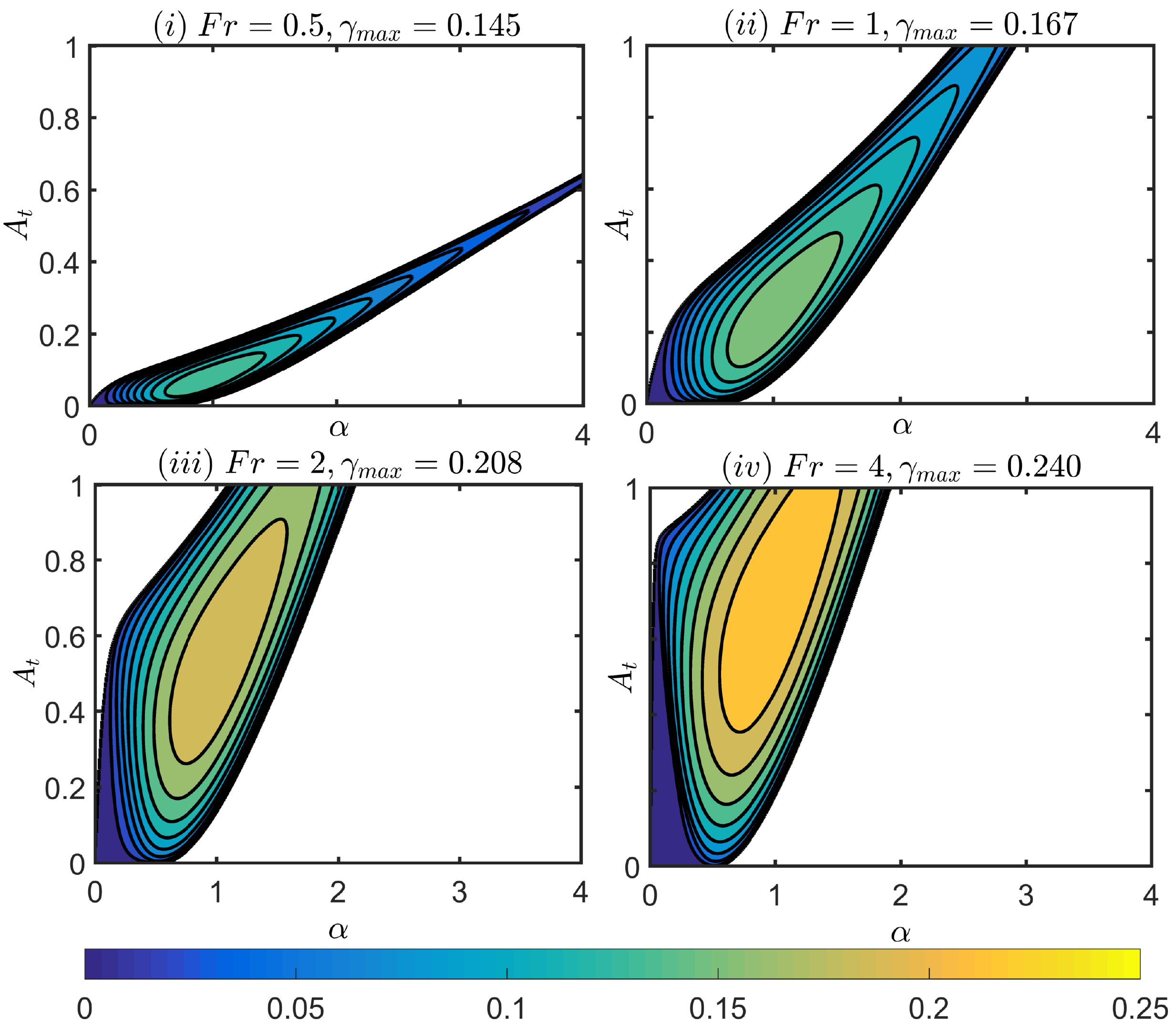}
   \caption{}
        \end{subfigure}%
        
		\begin{subfigure}[b]{\textwidth}
                \centering
                \includegraphics[width=100mm]{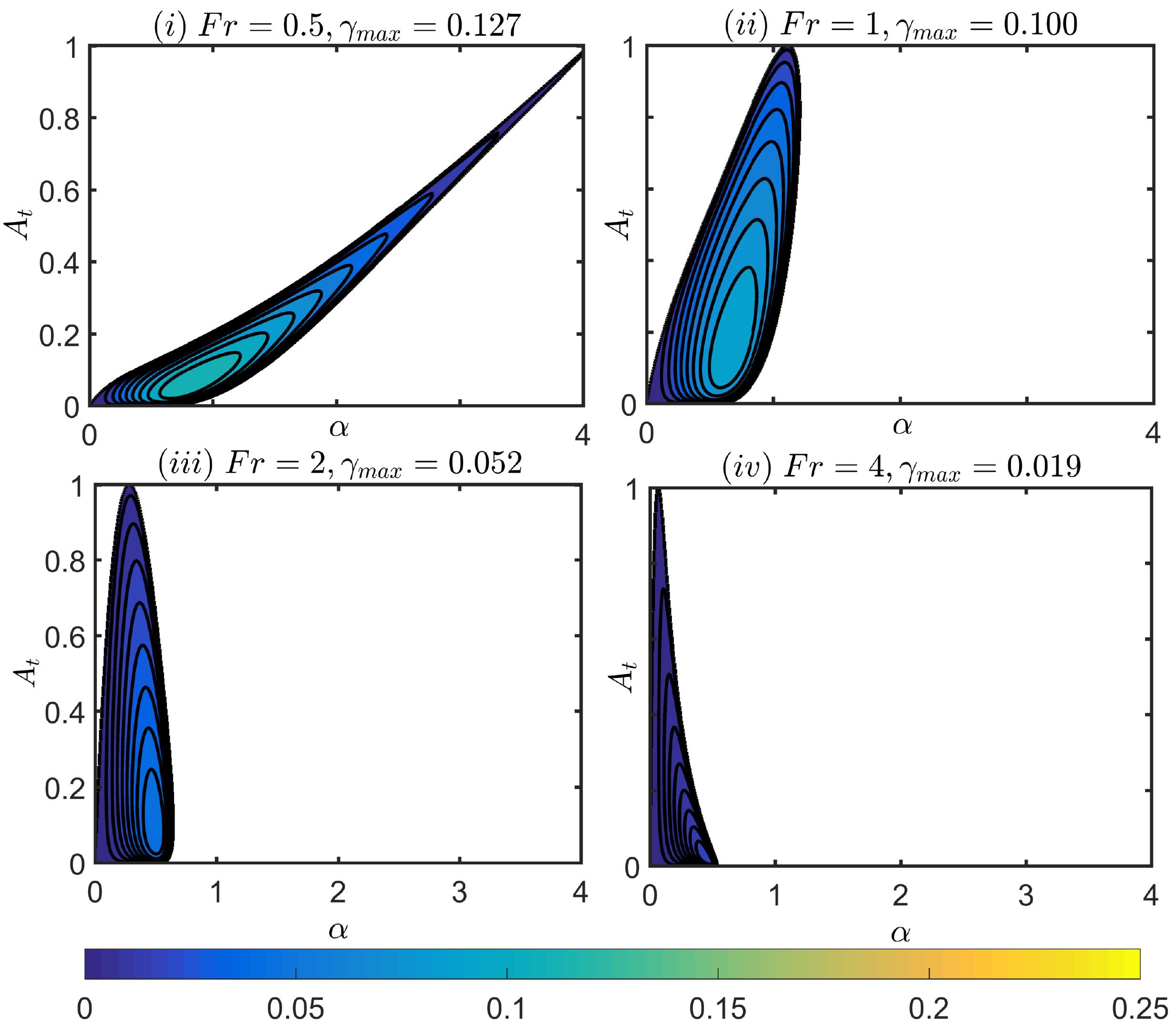}
                \caption{}
        \end{subfigure}%
        \caption{$A_t$--$\alpha$  plot showing growth rate contours for non-Boussinesq Holmboe configuration for different $Fr$ after (a) eliminating the wave $\mathcal{V^-}$ (b) eliminating the wave $\mathcal{V^+}$. The Froude number ($Fr$) and the maximum growth rate ($\gamma_{max}$)  is given in the title for each subplot.} 
        \label{fig:KH55}
\end{figure}

As can be seen from Fig.\ \ref{fig:KH55}(a),  an increase in $Fr$ after the removal of wave $\mathcal{V^-}$ leads to a gradual increase in the growth rate. However from Fig.\ \ref{fig:KH55}(b) it can be seen that in the absence of $\mathcal{V^+}$, increase in $Fr$ causes decrease in the maximum growth rate. Therefore, we conclude that interaction between the waves $\mathcal{SG^+}$ and $\mathcal{V^-}$ is enhanced by the shear while the interaction between the waves $\mathcal{SG^-}$ and $\mathcal{V^+}$ is suppressed by the shear.

From Fig.\ \ref{fig:Holmboe} it can also be observed that as $Fr$ increases, the contours tend to behave similar to the pure KHI case in the absence of  gravity (see Fig.\ \ref{fig:KH33}). As already mentioned in Sec. \ref{sec:KH},  this does not necessarily mean that the interaction is between the two vorticity waves i.e. $\mathcal{V^+}$ and $\mathcal{V^-}$.

 \section{Unstable stratification --   Rayleigh-Taylor instability in  presence of shear}
 \label{sec:RT_111}
 \begin{figure}
\centering\includegraphics[width=130mm]{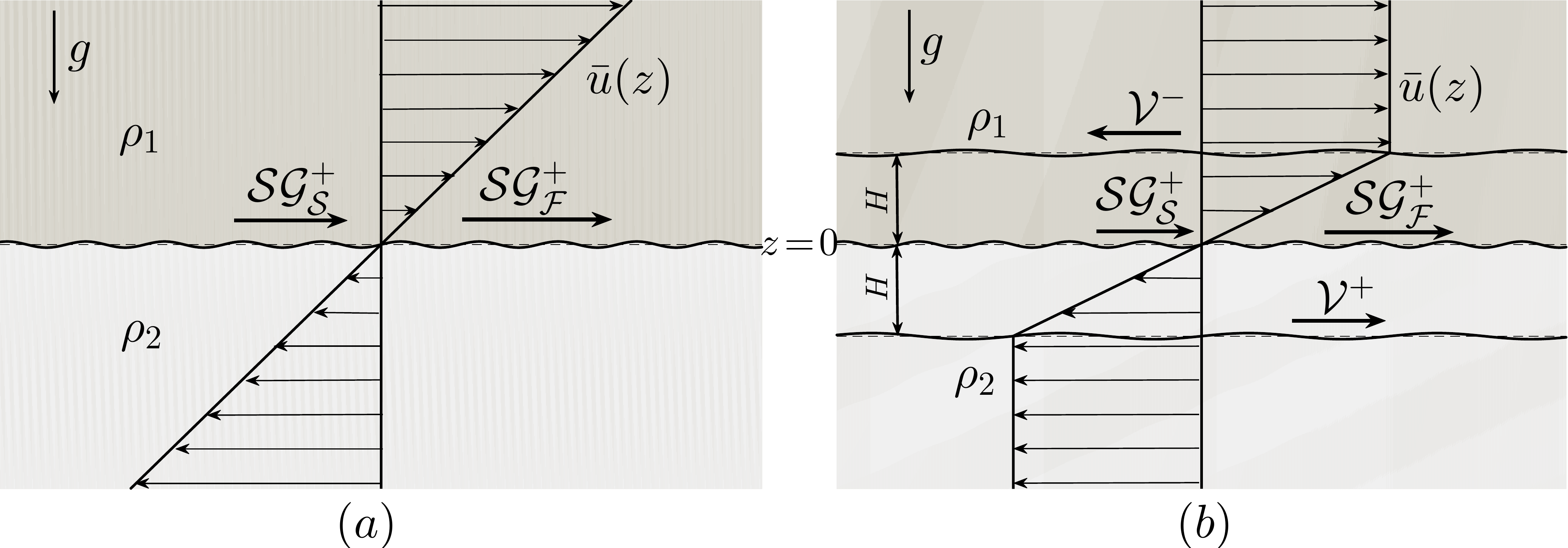}
  \caption{ Schematic of adversely stratified density interface (a) with a uniform shear (b) embedded in a shear layer forming a three interface system.}
  \label{fig:3_rti}
\end{figure} 
 
  Rayleigh--Taylor instability (RTI) is a familiar gravity driven flow phenomenon observed when a fluid of density $\rho_1$  rests on the top of a lighter one\citep{taylor1950instability} of density $\rho_2$ . In RTI with gravity $g>0$, the system is unstable for all Atwood number\footnote{Note that in this sections, we have defined Atwood number differently from the previous section so that RTI happens for $A_t>0$. Moreover, unlike shear instabilities, we will be non-dimensionalising the growth rates using $\sqrt{g/H}$ rather than $U/H$.}, $A_t \equiv (\rho_1-\rho_2)/(\rho_1+\rho_2)>0$ and for all wavenumber, $k$. 

The objective of this section is to study the effect of shear on a buoyancy interface with an adverse density stratification from the wave interactions perspective. We note here in passing that in the existing literature,  wave interactions perspective has mainly remained restricted to  stably stratified shear flows. {Previous studies on the effect of shear on RTI reveal contradictory results. \citet{kuo1963perturbations} showed that in presence of a uniform shear, RTI is inhibited. In another study by \citet{guzdar1982influence}, which used a continuous velocity profile for shear, it was also concluded that the RTI growth rate is inhibited by the application of shear. However, treating shear as a discrete velocity jump across the density interface, it was shown by \citet{zhang2005effect} that the presence of shear destabilizes the instability even further. \citet{benilov2002rti} argued that a uniform shear might stabilize RTI but any deviation from a uniform shear again `re-destabilizes' the flow.
To have a better understanding of the effect on shear on RTI, we make use of the broken line velocity profile which would help in underpinning the cause of the instabilities in terms of localized wave interactions.}  We initially assume nothing but a uniform shear across a buoyancy interface (Fig.\ \ref{fig:3_rti}(a)) and then we make the setting more realistic by assuming a finite shear layer (Fig.\ \ref{fig:3_rti}(b)). 
As discussed in the previous sections, when background velocity profile is of  shear layer type,  two more interfacial waves i.e. the vorticity waves $\mathcal{V^+}$ and $\mathcal{V^-}$ are introduced. These waves can interact to yield new instabilities, a possibility which we have also explored. 
 \subsection{RTI with a uniform shear}
 \label{sec:RTI_unif}
\begin{figure}
\centering
\includegraphics[width=120mm]{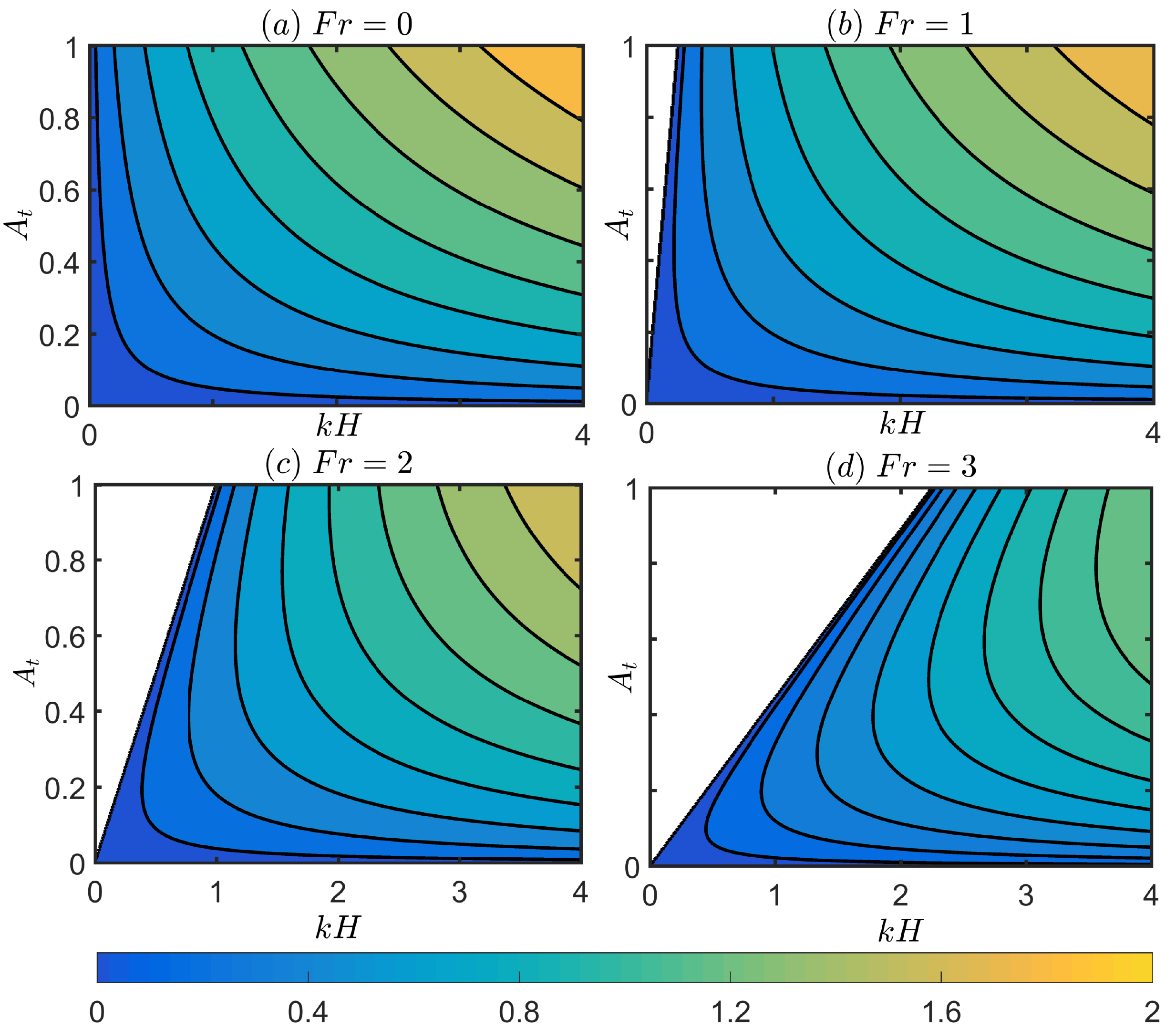}
  \caption{Stability diagram showing non-dimensional growth rate ${\gamma}/{\sqrt{g/H}}$ variation in the $A_t-kH$ plane for (a) $Fr=0$ (b) $Fr=1$ (c) $Fr=2$ (d) $Fr=3$. Here `H' is an arbitrarily chosen length scale.}
  \label{fig:211}
\end{figure}

In the presence of a uniform shear $Q$ across an adversely stratified buoyancy interface, the dispersion relation in the polynomial form is given by 
\begin{equation}\label{eq:dispersion11}
\omega^2-\left(\frac{\rho_1-\rho_2}{\rho_1+\rho_2}\right)Q\omega+gk\left(\frac{\rho_1-\rho_2}{\rho_1+\rho_2}\right)=0,
\end{equation}
from which we obtain (after using the definition of $A_t$)
\begin{align}
 \omega=\frac{A_tQ}{2}\pm\sqrt{\left(\frac{A_tQ}{2}\right)^2-A_tgk}.\label{Disp_Q2}
 \end{align}
Apart from the sign difference of $A_t$, the above equation is similar to \eqref{Disp_Q}. It is evident from the above relation that the presence of a uniform shear $Q$ always causes a suppression of the growth rate for a single interface system, and this effect is prominent at lower wavenumbers. The growth rates have been plotted in Fig.\ \ref{fig:211}. The value of shear (characterized by Froude number, $Fr\equiv Q/\sqrt{g/H})$ has been varied; it is observed that increasing the shear increases the stable region in the $A_t-kH$ {parameter space}. In the absence of any length scale, there is no physical significance of `$H$' and it can be chosen arbitrarily. 
{Perhaps the most important and non-intuitive result that we find exclusively when not making the Boussinesq approximation is the variation of growth rate with $A_t$. The growth rate no longer remains monotonic owing to the additional presence of $A_t$ in the shear term. This is because of the fact that the stabilizing effect of shear in RTI increases with an increase of density difference between the top and the bottom layer.} Thus, the instability, which is driven by the density difference in the `gravitational term', is also suppressed by the density difference in the `shear term'. Further, it can be easily shown that for a given `$k$' and $Q$, the maximum growth rate is obtained at $A_t=2gk/Q^2$, and further increasing the $A_t$ decreases the growth rate. 

It can be noted from  Eq.\ (\ref{eq:dispersion11}) that for $\rho_1>\rho_2$, the product of the roots  is positive; hence if both  of the roots are real, then they must be of the same sign. Physically this means that for sufficiently lower wavenumbers for which the system is stable, we obtain two stable waves propagating in the \emph{same direction}. For a positive $Q$, which is the case in this particular example, both of the waves (at lower wavenumbers) will be positively propagating. The slower of these waves is denoted by $\mathcal{SG_S^+}$ and the faster one is denoted by $\mathcal{SG_F^+}$. In this case, these two waves are shear--gravity waves. 
The waves obtained in this section should not be mistaken as interfacial gravity waves (which would have propagated in opposite directions) since stratification being adverse, gravity cannot provide the restoring force. Thus, the driving force for these waves has to be provided by the shear and the density jump while gravity only has a destabilizing role.

The uniform shear velocity profile (velocity having constant slope) considered here may be too simplistic for many realistic systems, where the characteristics of shear is usually captured by a `shear layer'. The latter will be focused in the next subsection. Nevertheless, velocity profile of constant slope is encountered in rotating inviscid fluids \citep{tao2013nonlinear}. The uniform shear set-up mathematically resembles that of   \citet{tao2013nonlinear} in many aspects, where the authors study the effect of rotation on RTI, the axis of rotation being normal to the direction of acceleration of the interface. The dynamic boundary condition in our case, i.e.\ \eqref{eq:DBC}, is similar to that obtained by   \citet{tao2013nonlinear} {in the context of rotating fluids} (see their Eq.\, (4)), where `$\Omega\psi$' like term appears due to Coriolis effects. Therefore, the  observation  in  \citet{tao2013nonlinear} that Coriolis force diminishes RTI growth rate is in line with the result that uniform shear inhibits RTI. 

\subsection{RTI within a shear layer}
\begin{figure}
\centering\includegraphics[width=120mm]{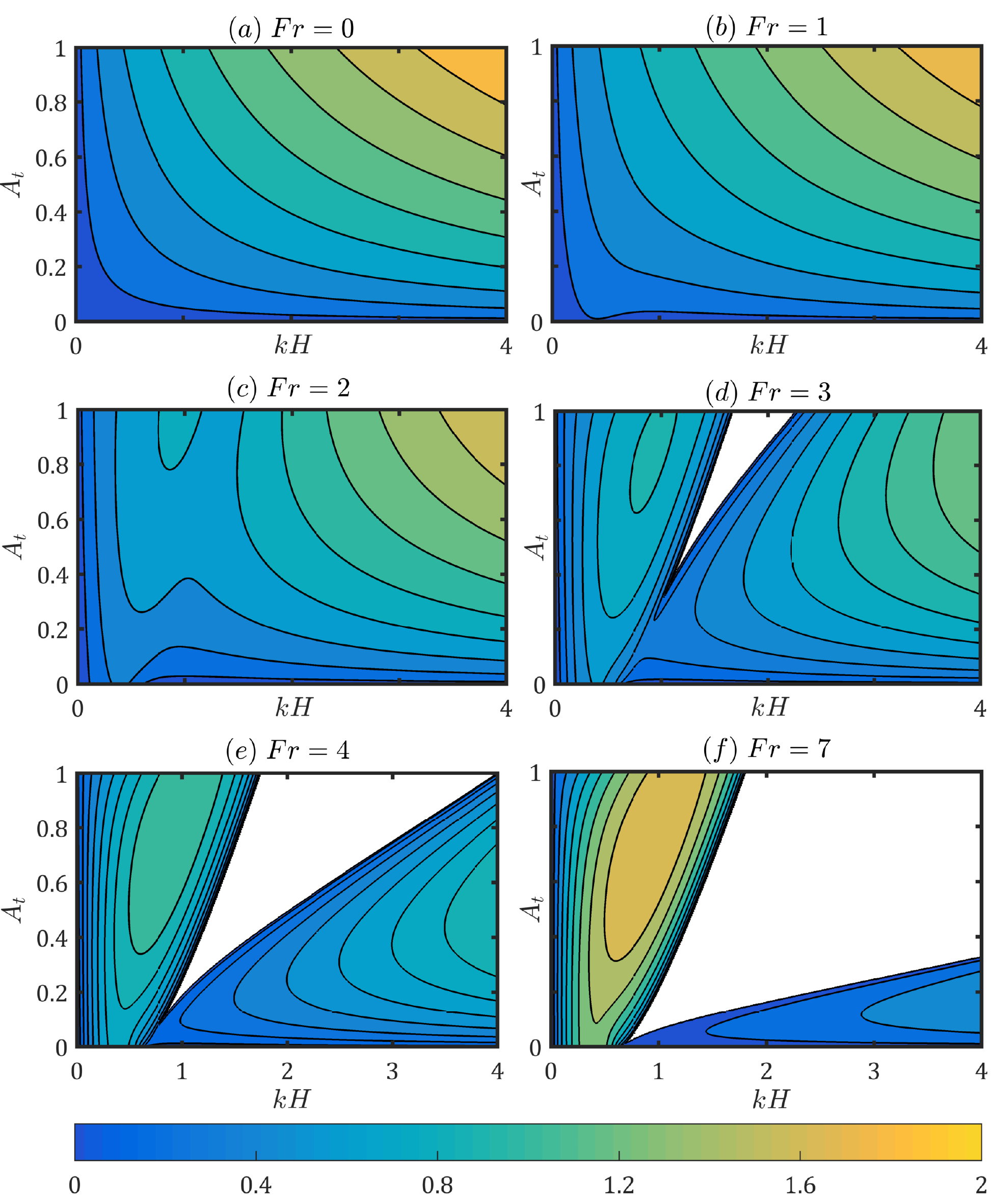}
  \caption{Stability diagram for RTI embedded in a shear layer. Two distinct instability regions are observed for high $Fr$, the right one being RTI while the left one is a shear instability.}
  \label{fig:RTI22}
\end{figure}
Here we extend the setting studied in the previous subsection  to have an unstable density interface sandwiched inside a `shear layer', see Fig.\  \ref{fig:3_rti}(b).
This set-up resembles the Holmboe instability set-up (which has already been discussed in Sec. \ref{sec:Holmb_in}), except that now the density interface has an unstable stratification. The system is described as follows:
\begin{equation}
\bar{\rho}(z) = \left\{
        \begin{array}{cc}
        \rho_1 & \quad 0 < z \\ 
        \rho_2 & \quad  z < 0
        \end{array}
    \right.
   \quad \bar{q}(z) = \left\{
        \begin{array}{cc}
        0& \quad H < z \\ 
        Q & \quad  -H < z<H \\ 
        0 & \quad z<-H.
        \end{array}
    \right.
\end{equation}
This system is inherently different from the Sec. \ref{sec:RTI_unif}. This is because here multiple interfaces are present; each interface can support one or more waves (Fig.\ \ref{fig:3_rti}(b)),  which implies that waves present at different interfaces can interact among themselves and lead to shear instability. Here, the waves $\mathcal{V^+}$ and $\mathcal{V^-}$ are vorticity waves whereas, waves `$\mathcal{SG_S^+}$' (the slower wave) and `$\mathcal{SG_F^+}$' (the faster wave),  are the shear--gravity waves.  Since there are 4 waves present in the system, the dispersion relation will naturally be a 4th order polynomial in $\omega$. To obtain this we proceed in the same way as we did in Sec. \ref{sec:gov_eq} - we introduce velocity potentials in each of the four regions and then use six kinematic and three dynamic boundary conditions. The dispersion relation in the non-dimensionalized form can be written as
\begin{equation}\label{eq:dispersion}
\tilde{\omega}^4+C_3\tilde{\omega}^3+C_2\tilde{\omega}^2+C_1\tilde{\omega}+C_0=0,
\end{equation}
Here $\;\tilde{\omega}\equiv\omega/(\sqrt{g/H})$ and
$$C_3=-2FrA_t\tanh{\alpha}(\tanh{\alpha}+1)^{-1};\quad C_2=Fr^2[\alpha-\alpha^2-\tanh{\alpha}(\tanh{\alpha}+1)^{-2}]+\alpha A_t;$$
$$C_1=Fr^3A_t[\alpha-\tanh{\alpha}(\tanh{\alpha}+1)^{-1}]^2;\quad C_0=-\alpha Fr^{-1}C_1;\,\,\,\,\,\,\,\,\,\,\,\,\,\,\,\,\,\,\,\,\,\,\,\,\,\,\,\,\,\,\,\,\,\,\,\,\,\,\,\,\,\,\,\,\,\,\,\,\,\,\,\,\,\,\,\,\,\,\,\,\,\,\,\,\,\,\,\,\,\,$$ where $\alpha\equiv kH.$ 
The non-dimensional growth rate $(\gamma/\sqrt{g/H})$ has been plotted w.r.t.\ $kH$ in Fig.\ \ref{fig:RTI22} for different values of $Fr$. Two distinct instability regions are observed for higher $Fr$ values - the right region of instability corresponds to RTI at $z=0$ due to the adverse buoyancy, and the left region is the shear instability region arising from two interacting pairs- (i) $\mathcal{V^-}$ and $\mathcal{V^+}$ (ii) $\mathcal{V^-}$ and $\mathcal{SG_F^+}$. This region of shear instability is very similar to that in Sec. \ref{sec:KH}, in which we showed that  these two waves form a counter-propagating configuration. Again, to conclusively separate the two instability modes in the set-up under consideration, we remove one of the vorticity waves at a time and see the consequence as we did in Sec. \ref{sec:KH}. First, we remove wave $\mathcal{V^-}$ by using the following density and shear profiles:     
\begin{equation}
\bar{\rho}(z) = \left\{
        \begin{array}{cc}
        \rho_1 & \quad 0 < z \\ 
        \rho_2 & \quad  z < 0
        \end{array}
    \right.
   \qquad \bar{q}(z) = \left\{
        \begin{array}{cc}
        Q & \quad H < z \\ 
        Q & \quad  -H < z<H \\ 
        0 & \quad z<-H
        \end{array}
    \right.
\end{equation}
\begin{figure}
\centering\includegraphics[width=130mm]{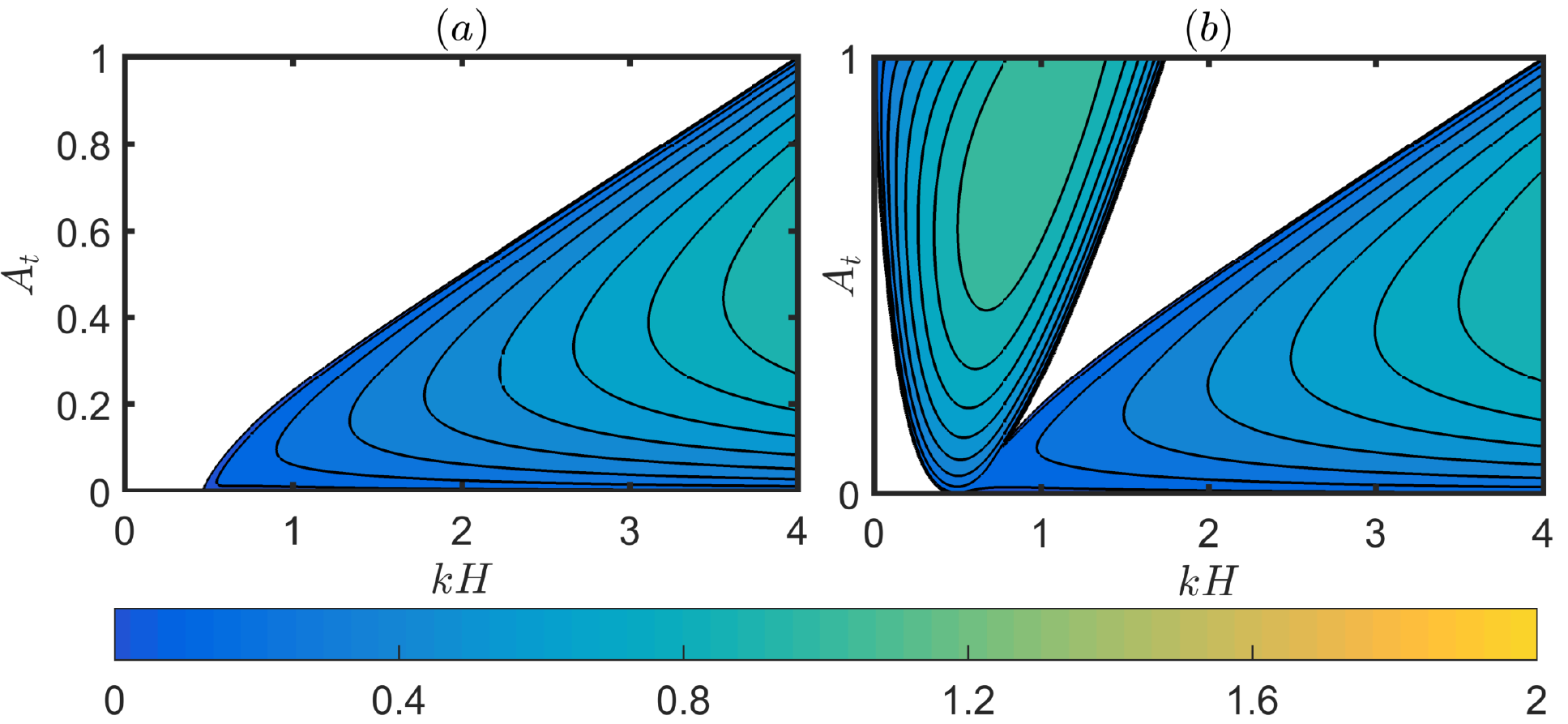}
  \caption{Stability diagram plot for $Fr=4$, (a) in the absence of wave $\mathcal{V^-}$ (b) in the absence of wave $\mathcal{V^+}$.}
  \label{fig:RTI3}
\end{figure}
Since the jump in base shear is eliminated at $z=H$, the vorticity wave $\mathcal{V^-}$ disappears. From Fig.\ \ref{fig:RTI3}(a) we see that removing wave $\mathcal{V^-}$  completely removes the region of shear instability. This shows that $\mathcal{V^-}$  plays a major role in the instability mechanism. Next, we eliminate wave $\mathcal{V^+}$ ( $\mathcal{V^-}$ being present) by eliminating the shear jump at $z=-H$, i.e.\ using the following profile:
\begin{equation}
\rho(z) = \left\{
        \begin{array}{cc}
        \rho_1 & \quad 0 < z \\ 
        \rho_2 & \quad  z < 0
        \end{array}
    \right.
   \qquad \bar{q}(z) = \left\{
        \begin{array}{cc}
        0 & \quad H < z \\ 
        Q & \quad  -H < z<H \\ 
        Q & \quad z<-H.
        \end{array}
    \right.
\end{equation}
The corresponding growth rate plot in Fig.\ \ref{fig:RTI3}(b) shows  very little difference with  Fig.\ \ref{fig:RTI22}(e), where wave $\mathcal{V^+}$ was present. The only small difference appearing at smaller values of $kH$ and at smaller $A_t$ is attributed to the presence of KHI in Fig.\ \ref{fig:RTI22}(e), which results from the interaction between waves $\mathcal{V^-}$ and $\mathcal{V^+}$. {We therefore conclude that a major role in such a shear instability is played by the waves $\mathcal{V^-}$ and $\mathcal{SG_F^+}$ rather than the two vorticity waves alone.}
{Further, we conclude that both  a uniform shear as well as a shear layer stabilize the Rayleigh--Taylor instability. However, presence of a shear layer allows the existence of two vorticity waves, which interact with a stable wave on the density interface and with themselves to give rise to a shear instability.  }


\section{Summary and Conclusions}
\label{sec:Sum_con}
 In this paper we have investigated the effects of density variation on both the inertial and gravity terms of the governing Navier-Stokes equations.
 To understand under what circumstances density variation can (or cannot) affect the inertial terms, we revisit the celebrated Boussinesq approximation. To this effect, instead of working with the momentum equations (which is a common practice), we have scrutinized the vorticity evolution equation,  Eq.\ (\ref{eq:2.3}). In the presence of moderate or large background shear, {the} Boussinesq approximation is applicable only to the flows with small density variations, i.e.
 \begin{equation*}
 A_t  \ll \mathcal{O}(1).
\end{equation*}
However, when the background shear is weak, the conditions for the validity of {the} Boussinesq approximation reads
$$\mathrm{either}\,\,\, A_t \ll \mathcal{O}\left(1\right)\,\, \mathrm{or}\,\,   Fr_c^2   \ll  \mathcal{O}(1),$$
where $Fr_c=c/\sqrt{gH}$ is a Froude number based on the ratio of the phase speed to the long wave speed. The second clause implies that a flow with $A_t=1$ can also be Boussinesq (provided shear is small), a finding contrary to the common notion that {the} Boussinesq approximation  is \emph{only} applicable to flows with small density variations. Our finding reveals that deep water surface gravity waves are Boussinesq; the Boussinesq approximation is valid even when background shear is present, provided it is weak. However, shallow water surface waves do not satisfy $Fr_c^2   \ll  \mathcal{O}(1)$, and hence are non-Boussinesq. These findings can have important consequences in modeling surface wave propagation as well as air-sea interactions.   
After broadening the applicability of the classical Boussinesq approximation, we have explored the existence of different kinds of interfacial waves that can be sustained in non-Boussinesq shear flows, and the ensuing instabilities owing to their resonant interactions.  Using potential flow theory we showed that a neutral wave can exist in the absence of a shear jump and without the need of gravity (the ``shear--density'' wave). {Similar waves were  also reported by \citet{behzad2014role}}. { The variation in $\bar{\rho}\bar{q}$ is shown to be a necessary and sufficient condition for the existence of such `non-gravity' waves in  stratified shear flows.} 

A {main} objective of this work has been to make a  proper distinction between \emph{density stratification} and \emph{buoyancy stratification}. Since Boussinesq flows are commonly studied,  density stratification and buoyancy stratification are often assumed synonymous. The distinction becomes clearer in non-Boussinesq flows, and our work is an attempt in this direction. In order to study the effect of an imposed shear layer on a density interface, three broad categories were formed: (A) neutrally stable density interface ($g=0$), (B) stably stratified density interface and (C) unstably stratified density interface. 
For case (A), we first studied a triangular jet profile (formed when a fluid of a given density 
flows through  a quiescent ambient fluid of different density). We understood this instability in terms of resonant interactions between shear--density and vorticity waves.   The most significant observation here was that increasing the density of the jet destabilizes the flow. Next we studied the Kelvin-Helmholtz instability and showed that in non-Boussinesq flows, it is basically the result of interactions between vorticity and shear--density waves. The interaction between two vorticity waves, as commonly known in homogeneous or Boussinesq flows, is only prominent at lower $A_t$. For the case (B), we studied non-Boussinesq Holmboe instability and showed that due to the non-Boussinesq effects growth of one of the Holmboe modes is hindered as shear is increased, while the growth rate of the other Holmboe mode increases. Ultimately, in the very high $Fr$ regime, the system is mainly governed by an instability between the shear--density wave and the vorticity wave traveling opposite to it. Finally for the case (C), we studied the effect of shear on Rayleigh-Taylor instability (RTI). We found that  not only shear stabilizes  RTI, but counter-intuitively, even increasing  $A_t$ might also stabilize RTI. Furthermore, if the shear is strong enough to stabilize RTI,  two neutrally propagating waves are obtained {on the density interface} which propagate in the same direction. The imposed shear, if part of a shear layer, can lead to a shear instability, which we show to be produced similar to the interaction between a vorticity wave and a shear--density wave.

\section*{Acknowledgements}
This work has been partially supported by the following grants: IITK/ME/2014338, STC/ME/2016176 and
ECR/2016/001493.

\appendix 
\section{Derivation of dynamic boundary condition in presence of piecewise background shear}
\label{app:A}
The inviscid Navier-Stokes equation within the bulk of a fluid of constant density $\rho$ is
\begin{equation}\label{eq:A1}
	\rho \left[ \frac{\partial \mathbf{u}}{\partial t}+\frac{1}{2} \nabla(\mathbf{u}\cdot\mathbf{u})- \mathbf{u}\times(\nabla\times \mathbf{u})\right]=-\nabla p -\nabla (\rho gz).
\end{equation}
Using the fact that there is no base vorticity generation in the bulk, we have $\nabla \times \mathbf{u}=\nabla \times \mathbf{\bar{u}}$=$Q\hat{j}$, where Q is  constant for each layer. Besides we use
\begin{equation}\label{eq:A2}
\mathbf{u}\times (Q\hat{j})=Q\nabla \psi=\nabla (Q\psi).
\end{equation}
Substituting \eqref{eq:A2} in \eqref{eq:A1} and removing the mean flow part, we are left with
\begin{equation}
	\rho \left[ \frac{\partial \mathbf{u}'}{\partial t}+\frac{1}{2} \nabla(\mathbf{u}'\cdot\mathbf{u}')+\nabla(\mathbf{\bar{u}}\cdot\mathbf{u}')- \nabla (Q\psi')\right]=-\nabla p' -\nabla (\rho g\eta).
\end{equation}
Since the perturbed flow is irrotational, we introduce $\mathbf{u}'=\nabla \phi'$. Moreover, since density is constant within each layer, we obtain
\begin{equation}
	\nabla\left[\rho \left( \frac{\partial \phi'}{\partial t}+\frac{1}{2} \nabla \phi'\cdot\nabla\phi'+\mathbf{\bar{u}}\cdot\nabla\phi'- Q\psi'+g\eta\right)+p'\right]=0.
\end{equation}
Since this is true for any arbitrary curve inside the domain, we have on integration
\begin{equation}
\rho \left( \frac{\partial \phi'}{\partial t}+\frac{1}{2} \nabla \phi'\cdot\nabla\phi'+\mathbf{\bar{u}}\cdot\nabla\phi'- Q\psi'+g\eta\right)+p'=c,
\end{equation}
where $c$ is an arbitrary function of time, which turns out to be zero in order to satisfy the unperturbed far-field condition. Thus, equating the pressure just above and just below the interface $z=\eta(x,t)$, at which the base flow velocity is $\mathbf{\bar{u}}=U\hat{i}$, and furthermore, dropping the nonlinear terms and  primes ($'$), we  obtain
\begin{equation}
\rho_1 \left( \frac{\partial \phi_1}{\partial t}+U\frac{\partial \phi_1}{\partial x}- Q_1\psi+g\eta\right)=\rho_2 \left( \frac{\partial \phi_2}{\partial t}+U\frac{\partial \phi_2}{\partial x}- Q_2\psi+g\eta\right).
\end{equation}

 \bibliography{paper1}
\end{document}